\documentclass[preprint,showpacs,superscriptaddress,amsmath,amssymb]{revtex4}
\usepackage{graphicx,color}

\begin{document}

\title{Stochastic models in population biology and their deterministic analogs}

\author{A. J. McKane}
\affiliation{Department of Theoretical Physics, University of Manchester, 
Manchester M13 9PL, UK}
\affiliation{Departments of Physics and Biology, University of Virginia, 
Charlottesville, VA 22904, USA}
\author{T. J. Newman}
\affiliation{Department of Physics and Astronomy, Arizona State University,
Tempe, AZ 85287, USA}
\affiliation{Departments of Physics and Biology, University of Virginia,
Charlottesville, VA 22904, USA}
 
\begin{abstract}
In this paper we introduce a class of stochastic population 
models based on ``patch dynamics.'' The size of the patch
may be varied, and this allows one 
to quantify the departures of these stochastic models from various 
mean field theories, which are generally valid as the patch size
becomes very large. These models may be used to formulate a broad
range of biological processes in both spatial and non-spatial
contexts. Here, we concentrate on two-species competition. We present
both a mathematical analysis of the patch model, in which we derive the
precise form of the competition mean field equations (and their first
order corrections in the non-spatial case), and simulation results. 
These mean field equations differ, in some important ways,
from those which are normally written
down on phenomenological grounds. Our general conclusion is that mean 
field theory is more robust for spatial models than for a single 
isolated patch. This is due to the dilution of stochastic effects in a 
spatial setting resulting from repeated rescue events mediated by 
inter-patch diffusion. However, discrete effects due to modest patch 
sizes lead to striking deviations from mean field theory even in a 
spatial setting.
\end{abstract} 


\maketitle
 
\section{Introduction} 
\label{intro}
 
Traditional theoretical ecology, in which the time evolution of population 
densities is described by differential equations, has a long history 
\cite{pie69,may74,bra01}. For a single species the simplest form of the 
governing equation is assumed to take the form $dN/dt = \Phi (N) N$, 
where $\Phi (N)$ describes the growth of the population. A common choice 
when modeling this growth is to take $\Phi (N) = r (1 - N/K)$, where $r$ 
and $K$ are two constants. By analogy, when describing the interaction of 
two species, it is natural to postulate that the populations of the two 
species, $N_{1}$ and $N_{2}$, change according to 
$dN_{1}/dt = f(N_{1}, N_{2})$ and $dN_{2}/dt = g(N_{1}, N_{2})$. 
The functions $f$ and $g$ are chosen according to whether the interactions are 
purely competitive, predator-prey like or include other effects such as 
co-operation. We will refer to descriptions of this kind as population-based; 
they are arrived at without the need for a detailed knowledge of the 
interaction between individuals and rely instead on assuming that the terms 
which arise in the governing equations represent the net effects of individual 
interactions in some generic way. Equations of this kind play such a central 
role in population biology, that many subsequent elaborations of the theory 
have taken them as the starting point: spatial variation is introduced by 
adding a drift term $\nabla^{2} N_{\alpha}$ ($\alpha=1,2$) to the right-hand 
side of the $\alpha$th equation, and the models are sometimes interpreted as 
referring to individuals by assuming that the functions $f$ and $g$ also 
describe interactions at the level of the individual.

In the last decade or so, an alternative approach to that of classical 
theoretical ecology described above has been developed. This involves 
abandoning the traditional population-level description in favor of an 
individual-based description in which explicit rules governing the 
interaction of individuals with each other and with the environment are given.
The popularity of these individual-based models (IBMs) is undoubtedly due to 
the continuing increase in the availability of powerful computers, but they 
also have other attractive features, such as the ability to directly model
individual attributes. At this point we should stress we are assuming that the
individuals of a given species in our model are identical, and thus the term 
IBM should not be confused with agent based models which are often designed
to study the ecological effects of behavioral and physiological variation 
among individuals. A better term might be ILM (individual level model), but
the term IBM has wide usage, and so we will use it here. In this paper we 
will be concerned with theoretical issues which relate to the connection 
between models defined at the individual level and those at the population 
level. Thus, the individuals in our models will be identical within a given 
species. The relation between population-level and individual-level 
descriptions has been a focus of discussion within the theoretical ecology 
community for some time \cite{der91,mcc93}. Some regard the nature of the 
population-based models as obvious and either write them down without comment, 
or derive mean-field equations by making an assumption of homogeneous mixing 
of the populations \cite{ran95,ber01}. However, there is also some recognition 
that the situation may be more complicated than this \cite{wil95,wil98}, and 
that the transition to a partial differential equation required for a spatial 
description, from the ordinary differential equation obtained by using 
mean-field theory, may not be as simple as just adding the term 
$\nabla^{2} N_{\alpha}$ \cite{dur94a,gan98,gan99}.

From a statistical physics perspective it is natural to expect that 
fluctuations may play an important role in these systems and lead to 
important differences between microscopic (IBM) and macroscopic 
(population-level) descriptions. A formal analysis of such issues has been 
presented for simple birth/death processes \cite{pel85}, and annihilation 
reactions \cite{car97} using the language of field theory and the 
renormalization group. Here, we take a different approach, using van Kampen's 
system size expansion, in order to probe the connection between more complex 
IBM's and mean field theories. This has some advantages, for instance it is 
not necessary using this technique to first construct a corresponding field 
theory, and then extract the mean-field equations and the Gaussian 
fluctuations about them to model the macroscopic behavior. There is also less
of an emphasis of the role of phase transitions, which are not usually of
prime interest in ecological models.  

We would like to stress from the outset that our use of 
``mean field theory'' or ``mean field model'' is in the spirit of
statistical mechanics. By ``mean field'' we mean the neglect of
correlations between degrees of freedom, allowing one to write the
mean of the product of two stochastic variables as the product of their
means. Thus, a spatial model in which statistical correlations are 
neglected is still a ``mean field model'' within this usage. This is in
contrast to some papers in population biology in which ``mean field'' is
reserved exclusively to describe non-spatial models.

In this paper, the models which we study will be defined only at the level of
direct interaction among individuals and in terms of local properties such as 
birth, death and migration rates. The population-based properties of the model 
will then be derived within a well-defined approximation scheme. We will show 
that, beginning with reasonable models at the individual level, the 
corresponding population-level models are similar in structure to those that 
we would naively write down, but have important differences. For instance, 
small scale diffusion may not simply translate into $\nabla^{2} N_{\alpha}$ 
terms and the parameters defined at the individual-based level may not map 
directly into their equivalents at the population level.

The individual interactions may be naturally introduced using a 
``patch model.'' From a biological perspective, a patch can be thought of as 
a small spatial region within which interactions between individuals occur. 
The patch is assumed to be sufficiently small that there are no spatial 
effects. In other words, there is complete mixing, and all individuals have
the same chance of potentially interacting with each other. In the 
non-spatial version of the model this simply amounts to stating that 
the probability of any individual dying in unit time should be proportional 
to the density of individuals existing at that time, and for processes which 
involve two individuals, the probability involved should be that found when 
drawing two of these types of individuals at random from a patch which 
contains all the individuals in the system. Not all constituents of the 
patch will correspond to individuals; some will correspond to empty sites in 
the spatial version of the model. A more detailed specification is given in 
Section \ref{basic}. Patch models such as this have been used in many areas 
of science \cite{fel68}, and often go under the name of ``urn models.'' Two 
early examples were the Ehrenfest urn, which was used to discuss the 
foundations of statistical mechanics, and the P\'olya urn, which was 
originally devised to describe contagious diseases \cite{joh77}. In both cases 
the problem of interest can be mapped onto an urn which contains balls of 
different colors, say black and white, which are drawn randomly one at a time. 
In the case of the Ehrenfest urn, each time a ball is drawn it is replaced by 
one of the other color. For the P\'olya urn, the chosen ball is replaced 
together with an extra ball of the same color. The relation to the modeling 
of contagious diseases should be clear: each occurrence of a particular color 
increases the probability of further occurrences. Since the introduction of 
these particular urn models, the notion has been generalized considerably. For 
example, both models fall into the class of urn models which have the 
properties that if a white ball is drawn, it is replaced together with $a$ 
white balls and $b$ black ones, and if a black ball is drawn, it is replaced 
together with $c$ white balls and $d$ black ones. A further generalization 
is to consider $r$ different colors, with the drawn ball of color $i$ being
replaced together with $a_{ij}$ balls of color $j\ $($i,j=1,\ldots,r$)
\cite{kot00}. It is also clear that two, or more, balls can be 
drawn together or, as we will have occasion to assume in this paper, two balls 
could be drawn for a fraction $\mu$ of the time and one ball for a fraction 
$(1-\mu)$ of the time.

Urn models are concrete realizations of stochastic processes with probabilities
which depend only on the instantaneous state of the system, that is, Markov
processes. They have proved useful in several areas of the biological 
sciences. Perhaps the most obvious application is in population genetics;
implicitly in the early work of Fisher and Wright \cite{cro70}, and 
explicitly in later developments 
\cite{hop84,don86,hop87,don96,dun97,fu98,sch01}. However they have also been 
used in a number of other areas such as the study 
of radioactive particles in animals \cite{ber77,she81,she83}, the study of 
patterns of vegetation \cite{svi99}, models of interaction between species 
\cite{sol00,mck00,sol02} and metapopulation models \cite{alo02}. The last 
three applications are closest to the ones discussed in this paper, but 
differ in that the balls represent forest or grassland in the first 
case, species in the second case and colonies in the third case, rather than 
individuals as in the present paper.

Since urn, or patch, models are representations of Markov processes,
the continuous time version of their dynamics may be described using a
master equation. The use of master equations is familiar to physicists, but
they are still not widely appreciated in the biological community (but see 
above references and Refs.~\cite{how01,zia02,zia04}, for instance). Once the 
process has been formulated in this way, we may use standard techniques to 
take the mean field limit and so obtain the corresponding population-based 
equation. Much of this paper will be taken up with a comparison of results 
obtained from the full model (averaged over many realizations) and the 
mean-field results. Our approach will be particularly useful in distinguishing 
situations where mean-field theory is a good approximation to the full theory 
and situations where it is not. We will mainly concentrate on competitive 
interactions in systems with one or two species, but our method applies 
equally well to predator-prey or epidemic models and multispecies communities.

The outline of the paper is as follows. We first motivate and define the 
rules for the non-spatial patch models in Section \ref{basic}, starting
with a single species model, and then generalizing to a two-species 
competition model. The full stochastic nature of the models is described
using master equations, and the associated mean field models are derived.
In the following section we go one step beyond mean field theory and derive
the dynamics of the Gaussian fluctuations about the mean field solutions. 
This is equivalent to not only following the mean position of the
probability distributions over time, but also describing their broadening.
In Section \ref{sim_nonspatial} we present numerical simulations of the
fully stochastic non-spatial models, for both one and two species. We
compare our results to numerical integration of both the mean field equations,
and the improved models with Gaussian fluctuations included. We find that
the agreement between mean field theory and simulations is excellent for
larger patches, as is to be expected, but that the breakdown of mean field
theory occurs precipitously below critical patch sizes which are still
fairly large. We proceed in Section \ref{spatial} to generalize our patch 
model formulation to spatially explicit population dynamics of two species 
competition. From the master equation we derive the spatial mean field 
equations. We note important differences between these equations and 
``intuitive'' versions which have appeared in the literature. These mean 
field equations are tested against simulations of the fully stochastic 
models in Section \ref{sim_spatial}. In this introductory paper we are 
unable to give a comprehensive analysis of the two species model. Instead 
we present two interesting scenarios, and note the successes and failures 
of the mean field equations, which are strongly dependent on patch size and
the presence/absence of interspecific competition. We end the paper
with our conclusions along with a discussion of future directions. Two
appendices contain technical details. The first concerns the system size 
expansion for two species and the second the formalism for spatial systems.
 
\section{Basic formalism and the non-spatial model} 
\label{basic}

In this section we will introduce the essential features of our approach by
formulating an individual based stochastic model of competition between two
species. We will then show that in the limit of large population sizes, the
time evolution of this model reduces to the well known differential equations  
describing the population growth of two competing species. 

The two species will be labeled $A$ and $B$. To motivate the approach
we will adopt, suppose that we model the interactions of $A$ and $B$
individuals in an area of land by subdividing it into $N$ plots of
equal area. The plot sizes are chosen so that each one either contains
one $A$ individual, or one $B$ individual, or neither an $A$ nor a
$B$. We will call the latter an empty site and label it by $E$. In a
spatial version of the model we would give rules for how $A, B$ and
$E$ interact, specify birth and death rates for $A$ and $B$, and allow
them to move to nearest neighbor sites. In later sections we will
describe such a model, but we will begin, for simplicity, by ignoring
the spatial aspects. We do this by imagining that we pluck all the $A,
B$ and $E$ from their particular sites and put them into a single
large patch, with no record of their original spatial locations. Any
memory of which individuals were nearest neighbors is now lost, and
any two individuals picked at random are just as likely to interact as
any other two similar individuals picked at random. In fact the time
evolution of the spatial model which would read: pick a site at
random, then pick a nearest neighbor of this site and implement the
interaction rule for the two chosen individuals, now reads: pick two
individuals from the patch and implement the reaction rule for the two
chosen individuals.

There are other, slightly different, ways of arriving at the above
picture.  For example, instead of dividing up the area of land into
sites containing either one individual or no individuals, it could be
divided up into a number of smaller patches, each of which contains several
individuals. In this way of thinking, each small patch
contains several $A, B$ and $E$ types, which interact with each other in the
same way as for the large patch described above, and which move by
exchange interactions with neighboring patches. While this defines a
slightly different spatial model, the non-spatial version of the model
is the same: all the individuals from the various patches are
collected together into a single large patch. Moreover, even the
spatial version of the model is in some sense a ``coarse-grained''
version of the original model --- several sites in that model when
viewed on a coarser scale can be reinterpreted as a site in the latter
model. These features will be explored in more detail when we discuss
the spatial aspects of these models in Section \ref{spatial}. For the
remainder of this section we will consider 
only the non-spatial model.

Suppose to begin with we consider the simpler case of a single
species, that is, a patch containing only $A$ and $E$ individuals. We
shall postulate that the population dynamics of the system can be
essentially described by three processes: birth, death, and
competition. The first and third processes will involve two
individuals: $AE \rightarrow AA$ (birth) and $AA \rightarrow AE$
(competition), but the second process involves only one individual: $A
\rightarrow E$ (death). These seem natural choices since, while death
can be modeled as constant, independent of the density of individuals,
the reduction in the numbers of $A$ due to competition and the growth
in numbers of $A$ due to births will be density dependent. In other
words, there will be a tendency for $AA$ to go to $AE$ because of
overcrowding, and for $AE$ to go to $AA$ due to the presence of
resources (space) to sustain a new individual.

The time evolution of the model can now be described. At each time
step we sample the patch. On a fraction $\mu$ of these occasions we
randomly choose 2 individuals and allow them to interact and for
$(1-\mu)$ of the draws we choose only one individual randomly. If in
the former case we draw two $E$ ``individuals'' or in the latter case
one $E$ ``individual'', we simply put them back into the patch. For all
other choices, an interaction may occur leading to the replacement of
a different set of individuals to those drawn. For each of these
processes we will introduce rate constants $b, c$ and $d$ as follows:
\begin{equation}
AA \stackrel{c}{\longrightarrow} AE , \ \ \ \ 
AE \stackrel{b}{\longrightarrow} AA , \ \ \ \
A \stackrel{d}{\longrightarrow} E\,.
\label{one_rates}
\end{equation}
We now only need to know the probabilities of drawing various combinations
from the patch. Simple combinatorics gives
\begin{eqnarray}
{\rm Probability\ of\ picking}\ AA & = & \mu\,\frac{n}{N}\,
\frac{(n-1)}{N-1}, \nonumber \\
{\rm Probability\ of\ picking}\ AE & = & 2\mu\,\frac{n}{N}\,
\frac{(N-n)}{N-1}, \nonumber \\
{\rm Probability\ of\ picking}\,\ A \,\ & = & (1-\mu)\,\frac{n}{N}\,,
\label{one_choose}
\end{eqnarray}
where the factor of 2 in the second term comes from the fact that the choices 
$AE$ and $EA$ are identical. These results enable us to write down
expressions for the transition probability, per unit time step, of the
system of individuals going from a state with $n$ $A$ individuals to a
state with $n'$ $A$ individuals. We denote this quantity by
$T(n'|n)$. Since only transitions from $n$ to $n \pm 1$ may take place
during one time step, the only non-zero $T(n'|n)$ are
\begin{eqnarray} 
T(n-1|n) & = & \mu\,c\,\frac{n}{N}\,\frac{(n-1)}{N-1} + 
(1-\mu)\,d\,\frac{n}{N}, \nonumber \\
T(n+1|n) & = & 2\mu\,b\,\frac{n}{N}\,\frac{(N-n)}{N-1}\,. 
\label{one_transprobs}
\end{eqnarray}

The process defined by (\ref{one_transprobs}) is a one-step Markov
process and so we can immediately write down a master equation
describing how the probability of having $n$ individuals present in
the patch, $P(n,t)$, changes with time \cite{van92}. The rate of
change of this quantity with time is simply the sum of transitions
from the states with $n+1$ and $n-1$ $A$ individuals to the state with
$n$ $A$ individuals, minus the sum of transitions from the state with
$A$ individuals to the state with $n+1$ and $n-1$ $A$ individuals:
\begin{eqnarray}
\frac{dP(n,t)}{dt} & = & T(n|n+1)P(n+1,t) + T(n|n-1)P(n-1,t) \nonumber \\ 
& - & T(n-1|n)P(n,t) - T(n+1|n)P(n,t)\,.
\label{one_master}
\end{eqnarray}
This set of coupled equations has to be solved subject to an initial
condition, typically $P(n,0) = \delta_{n,n_{0}}$, that is, a condition
stating that there are known to be $n_{0}$ individuals in the patch at
$t=0$. Care should also be taken with the boundary values $n=0$ and
$n=N$, since not all of the transitions are present in these
cases. From (\ref{one_transprobs}) we see that $T(-1|0)$ and
$T(N+1|N)$ are formally zero. So as long as we define
$T(0|-1)=T(N|N+1)=0$, thus we may use the general form
(\ref{one_master}) even for $n=0$ and $n=N$.

The master equation (\ref{one_master}) gives a complete description of the 
time evolution of the non-spatial model. In the next section we will discuss
the model predictions in more detail. Here we simply wish to make contact with
the mean field (i.e. the deterministic) version of the model, obtained by 
taking the $N \rightarrow \infty$ limit. This is most easily accomplished by 
multiplying the master equation (\ref{one_master}) by $n$ and summing over all
values of $n$. By shifting the variable in two of the sums on $n$ by $+1$ and 
$-1$, the following rate equation is obtained:
\begin{equation}
\frac{d\langle n \rangle}{dt} = \sum_{n=0}^{N}\,T(n+1|n)P(n,t) - 
\sum_{n=0}^{N}\,T(n-1|n)P(n,t)\,,
\label{one_rate_eqn}
\end{equation}
where angle-brackets signify averages over the possible states of the system.
Defining 
\begin{equation}
\tilde{c} = \frac{\mu c}{N-1}, \ \ \ \ \tilde{b} = \frac{\mu b}{N-1}, \ \ \ \  
\tilde{d} = \frac{(1-\mu)d}{N}\,,
\label{one_tilde}
\end{equation}
and using (\ref{one_transprobs}), (\ref{one_rate_eqn}) becomes
\begin{equation}
\frac{d\, }{dt}\,\frac{\langle n \rangle}{N} = 2 \tilde{b}\,
\left\langle \frac{n}{N} \left( 1 - \frac{n}{N} \right) \right\rangle -  
\tilde{c}\,\left\langle \frac{n}{N} \left( \frac{n}{N} - 
\frac{1}{N} \right) \right\rangle - 
\tilde{d}\,\left\langle \frac{n}{N} \right\rangle\,.  
\label{one_noverN}
\end{equation}

\smallskip

So far no approximation has been made in the derivation of (\ref{one_noverN}). 
However we now take the limit $N \rightarrow \infty$. In addition to 
eliminating the $1/N$ factor in the second term on the right hand-side of
(\ref{one_noverN}), it allows us to replace $\langle n^{2}\rangle$ by 
$\langle n \rangle^{2}$. This gives
\begin{equation}
\frac{d\phi}{dt} = 2 \tilde{b} \phi \left( 1 - \phi \right) - 
\tilde{c} \phi^{2} - \tilde{d} \phi\,, \ \ \ {\rm where} 
\ \ \phi \equiv \frac{\langle n \rangle}{N}\,.
\label{one_phi_eqn}
\end{equation}
Here $\phi(t)$ is the density of individuals in a given area. It is more
conventional to write this in the form
\begin{equation}
\frac{dN_{A}}{dt} = N_{A} \left( r - a N_{A} \right)\,, \ \ {\rm where}\
N_{A}(t) = \langle n \rangle = N \phi(t)\,,
\label{one_mft2}
\end{equation}
where 
\begin{equation}
r = 2 \tilde{b} - \tilde{d}, \ \ \ 
a = \frac{2 \tilde{b} + \tilde{c}}{N}\,.
\label{one_identi}
\end{equation}
Eq.~(\ref{one_mft2}) is the mean field equation of the model and is the 
familiar logistic equation, usually written down as a phenomenological 
description of the population growth of a single species with intraspecific 
competition. Here it is derived as the $N \rightarrow \infty$ limit of our 
stochastic model and thus provides a reasonable description of our system when 
the potential size (number of $A$ plus number of $E$ types) of the system is 
relatively large. Of course, this limit is purely formal. In practice what
we mean that if $N$ is of the order $10^4$, for instance, then this 
approximation is good if we are only interested in accuracies of up to 
$0.01\%$ (if the next order corrections are of order $1/N$) or $1\%$ (if the
next order corrections are of order $1/\sqrt{N}$). This approximation obviously
 cannot describe chance extinctions, which occur when $n$ is small, nor does 
it predict a mean time to extinction for the $A$ population. In regimes where 
these effects are important, it provides a poor description of the system, 
but this will inevitably be true of any purely deterministic description.

The necessity of introducing the ``empty site'' individuals $E$ should be
clear from the above derivation. In order to be able to define a
population density which changes with time we need a null population
which can be displaced if the $A$ species is successful and increase
if the $A$ population falters. It is also very natural to have a
ceiling on the growth of $A$ individuals ($N_{A} \leq N$) representing
a limit on the available resources. If no $E$'s were introduced in the
two species case, the two population sizes would not be independent
and would simply add up to $N$. The $A$ population would obey
(\ref{one_mft2}) and $N_{B}=N-N_{A}$. This is clear if we simply
imagine repeating the above derivation for a two species system by
replacing $E$ by $B$. Although the interaction rules would be altered
by this replacement, it would still be the case that
$N_{A}+N_{B}=N$. So it is vital to have ``empty space'' for
individuals to exploit in order to obtain a realistic population
dynamics.

We have discussed the single species case in some detail since the
construction of the master equation in the two species case (and, in
fact, in the $S$-species case for arbitrary $S$) follows similar
lines. We still draw 2 individuals $\mu$ of the time and 1 individual
$(1-\mu)$ of the time, but the processes and their rate constants are
now:
\begin{displaymath}
AA \stackrel{c_{11}}{\longrightarrow} AE , \ \  
AB \stackrel{c_{21}}{\longrightarrow} AE , \ \ 
BA \stackrel{c_{12}}{\longrightarrow} BE , \ \ 
A \stackrel{d_{1}}{\longrightarrow} E 
\end{displaymath}
\begin{equation}
BB \stackrel{c_{22}}{\longrightarrow} BE , \ \  
AE \stackrel{b_{1}}{\longrightarrow} AA , \ \ 
BE \stackrel{b_{2}}{\longrightarrow} BB , \ \ 
B \stackrel{d_{2}}{\longrightarrow} E\,.
\label{two_rates}
\end{equation}

The rate constants $c_{\alpha \alpha},\,\alpha=1,2$, represent intra-specific
competition and $c_{\alpha \beta}\ \alpha \neq \beta$, inter-specific 
competition. Analogous probabilities to (\ref{one_choose}) of choosing 
particular combinations of $A, B$ and $E$ are
\begin{eqnarray}
AA: \ \ \mu\,\frac{n}{N}\,\frac{(n-1)}{N-1}, \ & AE: & \ 2\mu\,\frac{n}{N}\,
\frac{(N-n-m)}{N-1}, \ \ A: \ \  (1-\mu)\,\frac{n}{N}, \nonumber \\
BB: \ \ \mu\,\frac{m}{N}\,\frac{(m-1)}{N-1}, \  & BE: \ & 2\mu\,\frac{m}{N}\,
\frac{(N-n-m)}{N-1}, \ \ B: \ \  (1-\mu)\,\frac{m}{N}, \nonumber \\
& AB: \ & 2\mu\,\frac{n}{N}\,\frac{m}{N-1}\,. 
\label{two_choose}
\end{eqnarray}

Transition probabilities now have initial and final states specified by two
integers. The transition probability per unit time from the state $(n,m)$ to 
the state $(n',m')$ will be denoted by $T(n', m'|n,m)$. The non-zero 
transition probabilities are
\begin{eqnarray} 
T(n-1,m|n,m) & = & \mu\,c_{11}\,\frac{n(n-1)}{N(N-1)} + 
(1-\mu)\,d_{1}\,\frac{n}{N} + 2\mu\,c_{12}\,\frac{mn}{N(N-1)}, \nonumber \\
T(n,m-1|n,m) & = & \mu\,c_{22}\,\frac{m(m-1)}{N(N-1)} + 
(1-\mu)\,d_{2}\,\frac{m}{N} + 2\mu\,c_{21}\,\frac{mn}{N(N-1)}, \nonumber \\
T(n+1,m|n,m) & = & 2\mu\,b_{1}\,\frac{n(N-n-m)}{N(N-1)}, \nonumber \\
T(n,m+1|n,m) & = & 2\mu\,b_{2}\,\frac{m(N-n-m)}{N(N-1)}\,.
\label{two_transprobs}
\end{eqnarray}

The master equation is an obvious generalization of (\ref{one_master}): 
\begin{eqnarray}
\frac{dP(n,m,t)}{dt} & = & T(n,m|n+1,m)P(n+1,m,t) + 
T(n,m|n-1,m)P(n-1,m,t) \nonumber \\
& + & T(n,m|n,m+1)P(n,m+1,t) + T(n,m|n,m-1)P(n,m-1,t) \nonumber \\
& - & \left\{ T(n-1,m|n,m) + T(n+1,m|n,m) + T(n,m-1|n,m) \right. \nonumber \\
& + & \left. T(n,m+1|n,m) \right\}\,P(n,m,t)\,.
\label{two_master}
\end{eqnarray}
This equation simply expresses the increase in $P(n,m.t)$ due to the four
possible transitions into the state $(n,m)$ described by 
$T(n,m|n\pm 1,m\pm 1)$, and the decrease due to transitions out of this state 
described by $T(n\pm 1,m\pm 1|n,m)$. The boundary and initial conditions are 
obvious analogs of those in the one-species case. The generalizations of 
(\ref{one_rate_eqn}) are quite simple, since none of the transition 
probabilities involving changes only in $m$ enter into the equation for 
$d\langle n \rangle/dt$, and none of the transition probabilities involving 
changes only in $n$ enter into the equation for $d\langle m \rangle/dt$:
\begin{eqnarray}
\frac{d\langle n \rangle}{dt} & = & \sum_{n,m=0}^{N}\,T(n+1,m|n,m)P(n,m,t) - 
\sum_{n,m=0}^{N}\,T(n-1,m|n,m)P(n,m,t), \nonumber \\
\frac{d\langle m \rangle}{dt} & = & \sum_{n,m=0}^{N}\,T(n,m+1|n,m)P(n,m,t) - 
\sum_{n,m=0}^{N}\,T(n,m-1|n,m)P(n,m,t)\,.
\label{two_rate_eqn}
\end{eqnarray}
We can now substitute the forms for (\ref{two_transprobs}) into this equation
and take the mean field limit $N \rightarrow \infty$. This allows us to factor
$\langle mn \rangle$ into $\langle m \rangle \langle n \rangle$, as well as
$\langle n^{2} \rangle$ into $\langle n \rangle^{2}$ and
$\langle m^{2} \rangle$ into $\langle m \rangle^{2}$ as before. The final 
result for the rate of change of population densities are the competition
equations 
\begin{eqnarray}
\frac{dN_{A}}{dt} & = & N_{A} \left( r_{1} - a_{11} N_{A} - 
a_{12} N_{B} \right) \ \ {\rm where} \ \ N_{A} = \langle n \rangle, 
\nonumber \\
\frac{dN_{B}}{dt} & = & N_{B} \left( r_{2} - a_{21} N_{A} - 
a_{22} N_{B} \right) \ \ {\rm and \ where} \ \ N_{B} = \langle m \rangle\,,
\label{two_mft2}
\end{eqnarray}
familiar from population biology textbooks. The parameters in (\ref{two_mft2}) 
are related to the parameters of the stochastic model by
\begin{equation}
r_{\alpha} = \frac{2\mu b_{\alpha}}{N-1} - \frac{(1-\mu)d_{\alpha}}{N}, \ \ \ \
a_{\alpha \alpha} = \frac{\mu(2b_{\alpha} + c_{\alpha \alpha})}{N(N-1)}, 
\ \ \ \ a_{\alpha \beta} = \frac{2\mu(b_{\alpha} + c_{\alpha \beta})}{N(N-1)}
\ (\alpha \neq \beta)\,.
\label{two_identi}
\end{equation}

As we will see in section \ref{spatial}, essentially the same kind of 
reasoning as that given above can be applied in the spatial version of the 
model. Before discussing this, however, we will investigate the large $N$ 
limit of (\ref{one_master}) and (\ref{two_master}) a little more carefully, 
obtaining corrections to mean field theory and comparing these to simulations.
 
\section{Beyond mean field theory for the non-spatial model} 
\label{nonspatial}

In the last section we gave arguments to show that the mean field versions
of the stochastic models we had introduced were indeed the deterministic
models conventionally used to describe these systems. In this section we 
will apply an elegant method due to van Kampen \cite{van92} which not only 
allows us to obtain these results in a more systematic way, but also gives a 
method of finding stochastic corrections to this deterministic result for 
large $N$. We will only work to next-to-leading order in this paper. This 
will give a Gaussian broadening to $P(n,t)$, or $P(n,m,t)$, with the peak of 
the distribution moving according to the relevant deterministic equation. We
will then compare these results with numerical simulations of the full 
stochastic process. The large $N$ expansion is very clearly explained by van 
Kampen in his book \cite{van92}, so we will content ourselves with 
giving a brief outline of the method as applied to the one species case. The 
two species calculation follows very similar lines.

We saw in the last section that, in the limit $N \rightarrow \infty$, the
variable $n$ became deterministic and equal to $N\phi(t)$. In this limit
the function $P(n,t)$ will be a delta-function. 
For large, but finite $N$, we would expect $P(n,t)$ to have a finite width of
order $N.N^{-1/2}=N^{1/2}$. Now $n$ is once again a stochastic variable, and
it is natural to bring out the large $N$ structure of the theory by 
transforming to a new stochastic variable $\xi$ by writing
\begin{equation}
n = N\phi(t) + N^{1/2} \xi\,.
\label{new_stoch} 
\end{equation}
We will not need to assume that the function $\phi(t)$ satisfies any particular
differential equation; if we simply choose it to follow the peak of the
distribution as it evolves in time, then the equation it satisfies will 
emerge. A new probability distribution function $\Pi$ is defined by 
$P(n,t)=\Pi(\xi,t)$, which implies that 
\begin{equation}
\dot{P} = \frac{\partial \Pi}{\partial t} - N^{1/2} \frac{d\phi}{dt}
\frac{\partial \Pi}{\partial \xi}\,.
\label{pdot}
\end{equation}
When using this formalism it is useful to rewrite the master equation 
(\ref{one_master}) using step operators which act on an arbitrary function of
$n$ according to ${\cal E}f(n)=f(n+1)$ and ${\cal E}^{-1}f(n)=f(n-1)$. This
gives
\begin{equation}
\frac{dP(n,t)}{dt} = \left( {\cal E} - 1 \right) \left[T(n-1|n)P(n,t)\right]
+ \left( {\cal E}^{-1} - 1 \right) \left[T(n+1|n)P(n,t)\right]\,.
\label{onestep_master}
\end{equation}
This form of the master equation is useful because the step operators have 
a simple expansion involving powers of the operator 
$N^{-1/2}\partial/\partial \xi$, which simplifies the identification of 
differing orders in $N^{-1/2}$. We shall assume the initial condition on the
equation to be $P(n,0) = \delta_{n,n_{0}}$.

Applying the method, and identifying powers of $N^{1/2}$, yields the 
macroscopic equation
\begin{equation}
\frac{d\phi}{dt} = \alpha_{1,0}(\phi)
\label{one_macro}
\end{equation}
to leading order and a Fokker-Planck equation
\begin{equation}
\frac{\partial \Pi}{\partial t} = -\alpha_{1,0}'(\phi)\frac{\partial }
{\partial \xi}\,\left[ \xi\Pi \right] + \frac{1}{2}\alpha_{2,0}(\phi)
\frac{\partial^{2} \Pi}{\partial \xi^{2}}
\label{one_FP}
\end{equation}
describing a linear stochastic process to next order. Here the functions 
$\alpha_{1,0}(\phi)$ and $\alpha_{2,0}(\phi)$ (we have used van Kampen's 
notation) are given by
\begin{eqnarray}
\alpha_{1,0}(\phi) & = & 2\tilde{b} \phi (1 - \phi) - \phi (\tilde{d} + 
\tilde{c}\phi) \nonumber \\
\alpha_{2,0}(\phi) & = & 2\tilde{b} \phi (1 - \phi) + \phi (\tilde{d} + 
\tilde{c}\phi)\,.
\label{one_alphas}
\end{eqnarray}

Since the Fokker-Planck equation (\ref{one_FP}) describes a linear process, 
its solution is a Gaussian. This means that the probability 
distribution $\Pi(\xi,t)$ is completely specified by the first two moments 
$\langle \xi \rangle_{t}$ and $\langle \xi^{2} \rangle_{t}$. Multiplying
(\ref{one_FP}) by $\xi$ and $\xi^{2}$ and integrating over all $\xi$ one finds
\begin{eqnarray}
\partial_{t}\langle \xi \rangle_{t} & = & \alpha_{1,0}'(\phi) \langle 
\xi \rangle_{t} \nonumber \\
\partial_{t}\langle \xi^{2} \rangle_{t} & = & 2\alpha_{1,0}'(\phi) \langle 
\xi^{2} \rangle_{t} + \alpha_{2,0}(\phi)\,.
\label{one_momenteqns}
\end{eqnarray}
The procedure is to solve (\ref{one_macro}) and obtain $\phi$ as a function
of $t$. This function is then substituted into (\ref{one_momenteqns}) and
these equations solved for $\langle \xi \rangle_{t}$ and 
$\langle \xi^{2} \rangle_{t}$. In the case of the first moment this may be 
performed quite generally to give
\begin{equation}
\langle \xi \rangle_{t} = \langle \xi \rangle_{0}\,\exp\left\{ \int^{t}_{0} 
d\tau \alpha_{1,0}'(\phi(\tau)) \right\}
\label{moment_solution}
\end{equation}
Choosing our initial condition to be
\begin{equation}
\phi(0) = \frac{n_{0}}{N}\,,
\label{one_IC}
\end{equation}
the initial fluctuations vanish, and $\langle \xi \rangle_{0}=0$. Thus
$\langle \xi \rangle_{t}=0$ for all $t$. 

This summarizes the method. We therefore start by solving (\ref{one_macro}), 
subject to (\ref{one_IC}). Defining
\begin{equation}
\rho \equiv 2\tilde{b} - \tilde{d} \ \ ; \ \ 
\sigma \equiv 2\tilde{b} + \tilde{c}\,,
\label{rho_and_sigma}
\end{equation}
for convenience, the general solution for $\rho \neq 0$ is
\begin{equation}
\phi(t) = \frac{\rho}{\sigma - Ae^{-\rho t}} , \ \ \ \ \rho \neq 0  
\label{macro_soln}
\end{equation}
where the constant $A$ is determined by the initial condition
\begin{equation}
A = \sigma - \frac{\rho}{\phi(0)} , \ \ \ \ \left( \phi(0) \neq 0 \right)
\label{A}
\end{equation}
If $\phi(0)=0$ then $\phi(t)=0$ for all $t$. If $\rho = 0$ a degenerate form 
of the solution exists and is given by 
\begin{equation}
\phi(t) = \frac{\phi(0)}{1 + \sigma \phi(0) t} , \ \ \ \ 
\left( \rho=0 \right)\,.
\label{zero_rho}
\end{equation}

This solution can now be substituted into the equation for 
$\langle \xi^{2} \rangle_{t}$ given in (\ref{one_momenteqns}). An integrating 
factor for this equation is $e^{2\rho t}/\phi^{4}(t)$, which yields upon use 
of the initial condition $\langle \xi^{2} \rangle_{0} = 0$ when $\rho \neq 0$, 
\begin{eqnarray}
\langle \xi^{2} \rangle_{t} & = & \frac{1}
{\left[ \sigma - Ae^{-\rho t} \right]^{4}}\,\left\{ 2\sigma^{2}\hat{b}(\hat{c}
+ \hat{d})\left[ 1 - e^{-2\rho t} \right] \right. \nonumber \\
& - & \sigma A \left[ 4\hat{b}^{2} + 10\hat{b}(\hat{c}+\hat{d}) + 
\hat{c}\hat{d} \right] e^{-\rho t} \left[ 1 - e^{-\rho t} \right] \nonumber \\
& + & \left. 2 A^{2} \rho \left[ 4\hat{b}^{2} + 4\hat{b}(\hat{c}+\hat{d}) + 
\hat{c}\hat{d} \right] t e^{-2\rho t} 
- A^{3}(2\hat{b}+\hat{d}) e^{-2\rho t} 
\left[ 1 - e^{-\rho t} \right] \right\}  
\label{second_moment}
\end{eqnarray}
Of course, care has to be taken when applying this approximation. If the 
distribution has significant interaction with the boundaries at $n=0$ or
$n=N$, the Gaussian approximation will break down. So, for example, if 
$\rho > 0$, the peak of the distribution will move from $n_0$ eventually
coming to rest at $N\rho/\sigma$. While this is happening the probability 
distribution broadens, eventually reaching its stationary value at
\begin{equation}
\lim_{t \rightarrow \infty}\langle \xi^{2} \rangle_{t} = 
\frac{2\tilde{b}(\tilde{c}+\tilde{d})}{(2\tilde{b}+\tilde{c})^{2}}\,.
\label{stationary}
\end{equation}
On the other hand, if $\rho \leq 0$, the peak of the distribution will 
eventually tend to zero, and so the Gaussian approximation will break down 
at some finite time. 

The case of two species follows in an exactly analogous manner: one
writes $n = N\phi(t) + N^{1/2}\xi$ and $m = N\psi(t) + N^{1/2}\eta$ and defines
a new probability distribution $\Pi$ via $P(n,m,t)=\Pi(\xi,\eta,t)$. The 
macroscopic equations obeyed by $\phi(t)$ and $\psi(t)$ are again found to be 
identical to those given in section \ref{basic}, and $\Pi$ is again found to 
satisfy a Fokker-Planck equation describing a linear process, and is therefore 
a multivariate Gaussian distribution. Details are given in Appendix A, where 
one sees that analytic expressions for the analogs of 
$\langle \xi^{2} \rangle_{t}$ cannot be obtained, in part at least, because the
macroscopic equations cannot be solved in closed form. However, we have solved 
them numerically, and we now go on to compare the large $N$ results in both 
the one species and two species cases with simulations of the original 
stochastic process.

\section{Simulations of non-spatial models} 
\label{sim_nonspatial}

In order to better understand the range of validity of mean-field
theory and its Gaussian corrections, we have performed numerical
simulations of the non-spatial model described above. The simulation
of stochastic models of population dynamics has progressed in line with the 
availability of high-speed computers over the last two decades. During
the beginning of this period, books on the stochastic dynamics of 
fluctuations in biological systems only mentioned simulations fleetingly 
\cite{gur82}, about a decade ago they had assumed a more central role 
\cite{ren91}, and are now regarded as essential in the understanding of 
these systems \cite{dur94b,dur99}. We refer the reader to these last two 
references for more details on how these simulations are carried out in 
practice.

In our numerical algorithm an ensemble of patches is iterated forward in
time. To achieve reasonable statistics we generally take the size
of the ensemble to be several thousand realizations.
In each small time step a single individual or a pair are selected from
each patch in turn and with the appropriate probabilities transitions
are made and the new individuals replaced.

We periodically average over the ensemble to measure both the mean
densities of individuals and the variance in the densities. Concurrent
with the stochastic simulation we also integrate forward the mean
field equations and the equations for the Gaussian variances 
(Eqs. (\ref{one_momenteqns}) and (\ref{av_x})-(\ref{av_xy})) to
allow a direct comparison. Forward integration of the differential
equations is performed using a second order Runge Kutta scheme.

We will present some examples of our numerical work which illustrate
the main effects that we have found. We have not performed an
exhaustive numerical analysis due to the large parameter space of the
competition models. The examples we give here are fairly typical
of a wide range of parameter space, and are chosen to illustrate a
variety of effects. Our results may be summarized by the statement
that so long as the size of the patch is large 
and the system is not close to extinction (such
that discrete effects play a strong role) then the mean field
equations, and the large $N$ correction yield a remarkably accurate 
description of the system dynamics. However, there is a fairly sharp
transition signaling the failure of mean field theory as the patch size 
is reduced below a critical value. For smaller patches, 
the quantitative precision of the mean field
theory fails badly. This inaccuracy gives way to qualitative errors
if one runs the system close to extinction. In this case the
probability distribution of the populations is poorly approximated by
a Gaussian function and one is compelled to abandon mean field theory
and its Gaussian corrections. It is significant that the critical
patch size depends sensitively on whether the patch contains one or two
species, and whether there is interspecific competition. To clarify
these statements we now lead the reader through some illustrative
examples. An exhaustive list of parameter values is given in Tables
1 and 2.

In Figure 1 we give an example of a moderately large patch 
($N=100$) 
containing a single species. The mean density soon settles down to a 
(quasi) steady-state value as does the variance in the population 
density. Mean field theory and the large $N$ corrections give very good
agreement. In Figure 2 we show results for an identical
situation but with the patch size reduced from 100 to 10. In this case
the quasi-steady-state is meaningless since extinction events are
frequent and the mean density (measured over the ensemble of patches)
steadily decays to zero \cite{gur82,new04}. Note, that the variance also 
decays to zero, since as time proceeds more and more realizations go 
extinct and the probability distribution of population densities is 
dominated by a delta function peak at zero. This illustrates the effect of
discrete individuals for small systems.

We now consider competition between two species in a single patch. We
only consider large patches, where naively, discrete individual effects
may be neglected. In Figure 3 we study the simplest case
in which the $A$ and $B$ individuals have identical birth, death, and
intraspecific competition rates. There is no explicit interspecific
competition (i.e. $c_{12}=c_{21}=0$), although the finiteness of the 
patch leads to indirect competition between all individuals
(see Eqs.~(\ref{two_mft2}) and (\ref{two_identi})). The patch is taken to 
be very large with a capacity of 400 individuals. We find 
satisfactory agreement between simulations and the mean field theory 
and its large-$N$ corrections. It is somewhat surprising that on reducing the 
patch size from 400 to 200 (Figure 4) the 
variance in the system increases steadily, far exceeding the 
large $N$ corrections. Returning to the large 400 capacity patch and
now introducing a small amount of interspecific competition
(Figure 5) we again find that the large $N$ corrections
fail to capture the growing fluctuations in the system. We conclude
from this, and other simulations, that mean field theory and its 
Gaussian corrections
can work well, but only for patches above a critical size.
This critical size is itself strongly dependent on the number of
species, various growth parameters, and the presence or not of
explicit interspecific competition.

\section{Spatial models} 
\label{spatial}

We have already discussed the spatial versions of the model in Section
\ref{basic}. In one version of the model, the area under consideration
is divided into a large number of patches, each containing a small
number of individuals, which are then identified with the sites of a
regular two-dimensional lattice (usually a square
lattice). Competition takes place between individuals of a particular
patch, and the birth rate is similarly only dependent on the
population density of the parental patch, but individuals are allowed
to migrate to nearest neighbor patches, if space is available (that
is, if an empty space, $E$, exists at the neighboring site).  In terms
of this picture, the
lattice consists of an array of patches, which interact through
migration of individuals from one patch to a nearest neighbor
patch. In the other version of the model introduced in section
\ref{basic}, each patch contains only one individual, thus the sites
of the lattice represent individuals rather than patches. In this case
competition and birth processes, as well as migration, depend on the
occupancy of nearest-neighbor sites. In both versions of the model,
the death rate is constant.

It is clear that many other variants are possible. In general, the first 
model will be more applicable to situations where individuals move on length
scales which are much larger than the communities they live in, and the 
second model more applicable when all these processes occur on scales which 
are of the same order. However, we will see that in the exploration of the 
nature of the mean-field limit and the importance of stochastic effects, which
is what interests us in this paper, these differences play a secondary role.  
In the next two sections we will follow the same program as was carried 
through for the non-spatial case: describing the stochastic process 
which defines the model, obtaining the mean-field limit, and finally
comparing the mean field equations with simulations of the full model. Much
of the mathematical detail will be relegated to Appendix B; unfortunately while
many of the ideas are simple generalizations of those introduced earlier, the
mathematical notation becomes (of necessity) rather complex and detracts from
the points which we wish to emphasize.

The simplest spatial process to describe is that of a single species
in the first version of the model. The processes in this case may be
broken down into 3 classes:
\begin{itemize}
\item[(i)]For a fraction $q_{1}$ of the events we randomly pick a site
$i$ and then randomly draw two individuals from within the patch at
that site. If two $E$ individuals are drawn, they are simply replaced,
otherwise the following interactions may occur
(c.f. Eqn. (\ref{one_rates}))
\begin{equation}
A_{i}A_{i} \stackrel{c}{\longrightarrow} A_{i}E_{i} , \ \ \ \ 
A_{i}E_{i} \stackrel{b}{\longrightarrow} A_{i}A_{i} . 
\label{one_rates_i}
\end{equation}
\item[(ii)]For a fraction $q_{2}$ of the events we randomly pick a
site $i$ and then randomly pick another site $j$ which is a nearest
neighbor of $i$. One individual is drawn from the patch at $i$ and another
from the patch at $j$.  If these two individuals are of the same type (two
$A$'s or two $E$'s) no action is taken, otherwise migration with a
rate constant $m$ may occur:
\begin{equation}
A_{i}E_{j} \stackrel{m}{\longrightarrow} E_{i}A_{j} , \ \ \ \ 
E_{i}A_{j} \stackrel{m}{\longrightarrow} A_{i}E_{j} . 
\label{one_mig}
\end{equation}
\item[(iii)]For a fraction $1 - q_{1} - q_{2}$ of the events we randomly pick 
a site $i$ and then randomly draw a individual from within the patch
at that site.  If an $E$ individual is drawn no action is taken,
otherwise death may occur at a constant rate $d$:
\begin{equation}
A_{i} \stackrel{d}{\longrightarrow} E_{i} . 
\label{one_death}
\end{equation}
\end{itemize}
The probabilities of choosing these various processes are in the case
of (i) and (iii) simply modifications of (\ref{one_choose}). The
modifications that are required are that all $n$ should be written as
$n_{i}$ to denote the number of $A$'s in the patch at site $i$, $\mu$
should be replaced by $q_1$, $(1-\mu)$ by $(1-q_{1}-q_{2})$ and all
terms multiplied by $\Omega^{-1}$, where $\Omega$ is the number of
sites in the lattice. We assume that the number of individuals in each
patch is the same for all sites and is denoted by $N$.  If the sites
$i$ and $j$ have already been chosen, the probabilities for the
processes (ii) are:
\begin{eqnarray}
{\rm Probability\ of\ picking}\ A_{i}E_{j} & = & q_{2}\,\frac{n_i}{N}\,
\frac{(N-n_j)}{N}, \nonumber \\
{\rm Probability\ of\ picking}\ E_{i}A_{j} & = & q_{2}\,\frac{n_j}{N}\,
\frac{(N-n_i)}{N}.
\label{one_mig_choose}
\end{eqnarray}

The transition probabilities, master equation, and the derivation of the 
population-level equation corresponding to this model, are discussed in 
Appendix B. The lattice version of this latter equation is given by 
(\ref{one_phieqn_spat}). On taking the continuum limit, defined by 
(\ref{continuum}), it becomes
\begin{equation}
\frac{\partial \phi}{\partial \tau} = \tilde{m} \nabla^{2} \phi + 
2 \tilde{b} \phi \left( 1 - \phi \right) - 
\tilde{c} \phi^{2} - \tilde{d} \phi \,,
\label{one_phi_eqn_spat}
\end{equation}
where $\phi({\bf x}, \tau)$ is the continuum version of 
$\langle n_{i}(\tau) \rangle/N$ in the limit $N \to \infty$ and where $\tau$ 
is a rescaled time. This equation is exactly the result (\ref{one_phi_eqn}) 
obtained in the non-spatial case, but with the addition of a 
$\nabla^{2} \phi$ drift term. We can write (\ref{one_phi_eqn_spat}) in a more 
standard form by defining a diffusion constant $D_{A} \equiv \tilde{m}$ and 
making the identification (\ref{one_identi}). This gives
\begin{equation}
\frac{\partial N_{A}}{\partial \tau} = D_{A} \nabla^{2} N_{A} + N_{A} 
\left( r - a N_{A} \right)\,, 
\label{one_eqn_spat}
\end{equation}
where, as before, $N_{A} \equiv N\phi$.

The discussion of the second version of the spatial model follows
similar lines. Now there is only one individual per site, and
therefore $n_{i}$ can only take on only two values: $0$ and $1$. In
addition, birth and competition processes, as well as migration,
depend on the occupancy of nearest neighbor sites. Therefore there are
only 2 classes of processes:
\begin{itemize}
\item[(i)]For a fraction $\mu$ of the events we randomly pick a site
$i$ and then randomly pick another site $j$ which is a nearest neighbor of 
$i$. If these sites are both $E$'s no action is taken, otherwise migration 
with a rate constant $m$ may occur according to (\ref{one_mig}), or 
birth and competition may take place with rate constants $b$ and $c/2$ 
respectively:
\begin{equation}
A_{i}E_{j} \stackrel{b}{\longrightarrow} A_{i}A_{j} , \ \ 
E_{i}A_{j} \stackrel{b}{\longrightarrow} A_{i}A_{j} , \ \
A_{i}A_{j} \stackrel{c/2}{\longrightarrow} E_{i}A_{j} , \ \  
A_{i}A_{j} \stackrel{c/2}{\longrightarrow} A_{i}E_{j} . 
\label{one_bcm}
\end{equation}
The factor of $1/2$ has been introduced into the rate constant for 
competition in order be consistent with the non-spatial case and the first 
version of the spatial case. 
\item[(ii)]For a fraction $1 - \mu$ of the events we randomly pick 
a site $i$. If the site contains an $E$ individual, no action is taken. 
Otherwise death may occur at a constant rate $d$ given by (\ref{one_death}).
\end{itemize}
Since there is only one individual per each site, the probabilities of picking
$A_{i}E_{j}$ or $E_{i}A_{j}$ are simply (\ref{one_mig_choose}) with $N=1$
and $q_{2}$ replaced by $\mu$. Similarly, the probability of picking $A_{i}$ 
is as in the first version, but again with $N=1$ and with $(1-q_{1}-q_{2})$ 
replaced by $(1-\mu)$. The only new feature is:
\begin{equation}
{\rm Probability\ of\ picking}\ A_{i}A_{j} = \mu\,n_{i} n_{j}\,.
\label{one_ver2_comp}
\end{equation}
Just as before, we have assumed that the sites $i$ and $j$ were
already chosen, so that the above probabilities only represent the
choices of types of individuals at these chosen sites/patches and also
the choice of the number of individuals in an event (one or two). In
the first model, we denoted the number of sites in the lattice by
$\Omega$. This was independent of $N$, the number of individuals in a
patch. It was this latter quantity that we allowed to become
infinitely large, in order to deduce the population-level
description. In this second version, there is only one individual per
site, and so it is the number of lattice sites (now denoted by $N$)
which we take to be infinitely large.  More details of this approach
are given in Appendix B, where it is shown that, in the large $N$
limit, the population-level description is again given by
(\ref{one_phi_eqn_spat}) --- albeit with slightly different
definitions of the parameters. This is not a surprise; we would expect
there to be a large number of IBMs which differ in detail, but which
have the correct qualitative features, and give the same
population-level description.

The partial differential equation (\ref{one_eqn_spat}) is simply the 
ordinary differential equation for the non-spatial case (\ref{one_phi_eqn}),
but with a term $\nabla^{2} N_{A}$ added. So the corresponding spatial 
description is indeed obtained by using the simplest prescription. However, 
this will not turn out to be the case when more than one species are present. 
It is this scenario which is of most interest to us in this paper; we have 
described the one species case in some detail, largely because it is 
technically simpler, and therefore the crucial steps in the argument clearer.
The many species case may differ at the population-level, but the the 
setting up of the IBMs and the derivation of the population-level equations
are a straightforward generalization of the one-species case.

Let us once again begin with the first version of the model where
birth, competition, and death processes are purely local (they take
place in a single patch at a specific site on the lattice) and only
the process of migration involves nearest-neighbor patches. All of the
transitions are variants of those in models previously considered in
this paper. Specifically the 3 classes of processes are:
\begin{itemize}
\item[(i)]For a fraction $q_{1}$ of the events we randomly pick a site
$i$ and then randomly draw two individuals from within the patch at that
site. If two $E$ individuals are drawn, they are simply replaced, otherwise
the interactions are given by the 6 two individual interactions in
(\ref{two_rates}) with a site index $i$ added on to the $A$, $B$ and
$E$ individuals.
\item[(ii)]For a fraction $q_{2}$ of the events we randomly pick a
site $i$ and then randomly pick another site $j$ which is a nearest
neighbor of $i$. One individual is drawn from the patch at $i$ and another
from the patch at $j$.  If neither of these two individuals are $E$'s (no
space) or both are $E$'s (no migration possible), then no action is
taken, otherwise migration with rate constants $m_{1}$ or $m_{2}$ may
occur:
\begin{equation}
A_{i}E_{j} \stackrel{m_{1}}{\longrightarrow} E_{i}A_{j} , \ \  
E_{i}A_{j} \stackrel{m_{1}}{\longrightarrow} A_{i}E_{j} , \ \
B_{i}E_{j} \stackrel{m_{2}}{\longrightarrow} E_{i}B_{j} , \ \  
E_{i}B_{j} \stackrel{m_{2}}{\longrightarrow} B_{i}E_{j} .
\label{two_mig}
\end{equation}
\item[(iii)]For a fraction $1 - q_{1} - q_{2}$ of the events we randomly pick 
a site $i$ and then randomly draw a individual from within the patch
at that site.  If an $E$ individual is drawn no action is taken,
otherwise death may occur at constant rates $d_{1}$ or $d_{2}$:
\begin{equation}
A_{i} \stackrel{d_{1}}{\longrightarrow} E_{i} , \ \ \ \ 
B_{i} \stackrel{d_{2}}{\longrightarrow} E_{i} . 
\label{two_death}
\end{equation}
\end{itemize}
The probabilities of choosing these various processes are in the case of (i) 
and (iii) simply modifications of (\ref{two_choose}). The modifications that 
are required are exactly those we described in the similar version of the 
one species case: the $n$'s and $m$'s should be written as $n_{i}$ and $m_{i}$
respectively, $\mu$ should be replaced by $q_1$, $(1-\mu)$ by $(1-q_{1}-q_{2})$
and all terms multiplied by $\Omega^{-1}$. The migration of $A$'s and $B$'s 
are independent of each other, and so are described in exactly the same way as
for single species. Details are given in Appendix B where it is shown that,
after the continuum limit has been taken, the equations for
\begin{equation}
\phi({\bf x}, \tau) \equiv \lim_{N \to \infty} 
\frac{\langle n_{i} (\tau) \rangle}{N} \ \ \ {\rm and} \ \ \  
\psi({\bf x}, \tau) \equiv \lim_{N \to \infty} 
\frac{\langle m_{i} (\tau) \rangle}{N}
\label{phi_psi_defn}
\end{equation}
are
\begin{eqnarray}
\frac{\partial \phi}{\partial \tau} &=& \tilde{m}_{1} \nabla^{2} \phi + 
\tilde{m}_{1} \left( \phi \nabla^{2} \psi - \psi \nabla^{2} \phi \right)
\nonumber \\
&+& 2 \tilde{b}_{1} \phi \left( 1 - \phi - \psi \right) - 
\tilde{c}_{11} \phi^{2} + 2\tilde{c}_{12} \phi \psi - \tilde{d}_{1} \phi \,,
\label{two_phi_eqn_spat} 
\end{eqnarray}
and
\begin{eqnarray}
\frac{\partial \psi}{\partial \tau} &=& \tilde{m}_{2} \nabla^{2} \psi + 
\tilde{m}_{2} \left( \psi \nabla^{2} \phi - \phi \nabla^{2} \psi \right)
\nonumber \\
&+& 2 \tilde{b}_{2} \psi \left( 1 - \phi - \psi \right) - 
\tilde{c}_{22} \psi^{2} + 2\tilde{c}_{21} \psi \phi - \tilde{d}_{2} \psi \,.
\label{two_psi_eqn_spat}
\end{eqnarray}

The second version of the two-species model has $n_{i}=0, 1$ and $m_{i}=0, 1$,
with birth, competition, and migration depending on nearest-neighbor 
occupancies. The 2 classes of processes are
\begin{itemize}
\item[(i)]For a fraction $\mu$ of the events we randomly pick a site
$i$ and then randomly pick another site $j$ which is a nearest neighbor of 
$i$. If these sites are both $E$'s no action is taken, otherwise migration 
may occur according to (\ref{two_mig}), birth according to
\begin{equation}
A_{i}E_{j} \stackrel{b_{1}}{\longrightarrow} A_{i}A_{j} , \ \ 
E_{i}A_{j} \stackrel{b_{1}}{\longrightarrow} A_{i}A_{j} , \ \
B_{i}E_{j} \stackrel{b_{2}}{\longrightarrow} B_{i}B_{j} , \ \  
E_{i}B_{j} \stackrel{b_{2}}{\longrightarrow} B_{i}B_{j} , 
\label{two_b}
\end{equation}
and competition according to
\begin{displaymath}
A_{i}A_{j} \stackrel{c_{11}}{\longrightarrow} A_{i}E_{j} , \ \ 
B_{i}A_{j} \stackrel{c_{12}}{\longrightarrow} B_{i}E_{j} , \ \
A_{i}B_{j} \stackrel{c_{21}}{\longrightarrow} A_{i}E_{j} , \ \  
B_{i}B_{j} \stackrel{c_{22}}{\longrightarrow} B_{i}E_{j} , 
\end{displaymath}
\begin{equation}
A_{i}A_{j} \stackrel{c_{11}}{\longrightarrow} E_{i}A_{j} , \ \ 
A_{i}B_{j} \stackrel{c_{12}}{\longrightarrow} E_{i}B_{j} , \ \
B_{i}A_{j} \stackrel{c_{21}}{\longrightarrow} E_{i}A_{j} , \ \  
B_{i}B_{j} \stackrel{c_{22}}{\longrightarrow} E_{i}B_{j} . 
\label{two_m}
\end{equation}
\item[(ii)]For a fraction $1 - \mu$ of the events we randomly pick a
site $i$. If the site contains an $E$ individual, no action is taken.
Otherwise death may occur according to (\ref{two_death}).
\end{itemize}
The probabilities of picking two individuals, one of which is an $E$,
are the same as in the first version of the model, but with $N=1$ and
$q_{2}$ replaced by $\mu$. The probabilities of picking a single
individual are similarly related to those found in the first
version. The probabilities associated with picking $A_{i}A_{j}$ is
given by (\ref{one_ver2_comp}), $B_{i}B_{j}$ by $\mu m_{i} m_{j}$ and
$A_{i}B_{j}$ by $\mu n_{i} m_{j}$. In Appendix B we describe how, in
the limit where the number of lattice sites $N$ becomes infinitely
large, the continuum versions of $\langle n_{i} \rangle$ and $\langle
m_{i} \rangle$ again satisfy equations (\ref{two_phi_eqn_spat}) and
(\ref{two_psi_eqn_spat}).

So, in summary, both versions of the IBMs we have discussed in this paper 
give rise to the same population-level equations. This is true whether there 
is only a single species in the system, or whether two species are present.  
In the one species case this equation is given in standard form by 
(\ref{one_eqn_spat}). To write the two species equations 
(\ref{two_phi_eqn_spat}) and (\ref{two_psi_eqn_spat}) in standard form we 
make the identification (\ref{two_identi}) and introduce diffusion constants
\begin{equation}
D_{A} = \tilde{m}_{1}, \ \ 
D_{B} = \tilde{m}_{2}, \ \ 
D_{1} = \frac{\tilde{m}_{1}}{N}, \ \ 
D_{2} = \frac{\tilde{m}_{2}}{N}.
\label{diffusion_consts}
\end{equation}
This gives
\begin{eqnarray}
\frac{\partial N_{A}}{\partial \tau} &=& D_{A} \nabla^{2} N_{A} 
+ D_{1} \left( N_{A} \nabla^{2} N_{B} - N_{B} \nabla^{2} N_{A} \right) 
+ N_{A} \left( r_{1} - a_{11} N_{A} - a_{12} N_{B} \right), 
\label{two_eqnA_spat} \\
\frac{\partial N_{B}}{\partial \tau} &=& D_{B} \nabla^{2} N_{B} 
+ D_{2} \left( N_{B} \nabla^{2} N_{A} - N_{A} \nabla^{2} N_{B} \right)
+ N_{B} \left( r_{2} - a_{21} N_{A} - a_{22} N_{B} \right), 
\label{two_eqnB_spat}
\end{eqnarray}
where $N_{A} = N\phi$ and $N_{B} = N\psi$. Unlike (\ref{one_eqn_spat}), these 
are {\em not} the standard equations found in population biology textbooks.

The additional terms which appear in (\ref{two_eqnA_spat}) and 
(\ref{two_eqnB_spat}), but not in the standard equations, are antisymmetric 
in $N_{A}$ and $N_{B}$ and involve derivatives, and so do not appear in 
non-spatial models or spatial models with only one species. Their structure 
is dictated by the way that migration is modeled at the individual level. 
Since their occurrence is generic, they will also appear in spatial models 
derived from IBMs having three or more species. Although these terms have not 
to our knowledge been discussed in the context of ecological models, they are 
well-known in the context of interspecies diffusion~\cite{sch92,kor97} in 
physics, and they also appear in quantum field theory~\cite{gas66}.

\section{Simulations of spatial models} 
\label{sim_spatial}

The simulations of spatial competition was performed in a directly
analogous fashion to the non-spatial model described in Section
\ref{sim_nonspatial}. We confined ourselves to one spatial dimension
for simplicity. In addition, we only simulated the version of
the model where a patch of size $N$ was placed at each site of the
one-dimensional lattice of size $L$. Within each patch the usual dynamics of
competition is played out. Furthermore, in each small time step there
is a small probability of dispersal of individuals from a given patch
to the two neighboring patches, just as encoded in the master
equation. We generally started with initial conditions in which
species $A$ and $B$ were spatially separated and then proceed to
intermix and compete as individuals diffuse from patch to patch.  The
mean field equations (\ref{two_eqnA_spat}) and (\ref{two_eqnB_spat})
are again integrated forwards in time using second-order Runge-Kutta
methods. An exhaustive list of parameter values is given in Table 3.

In a spatial system such as this, extinction is much less of a problem
since should a patch become empty it will soon be restocked from
neighboring patches. Despite the weakened effect of discreteness in
small patches, we still find that the behavior of the spatial systems
differs significantly from mean field theory when patch sizes are
below a critical value (of approximate value 50 for the results
presented here).  As before, we have chosen to present two typical
scenarios.

In Figure 6  we show the early and late time behavior for a
system in which initially the $A$ individuals occupy the left half of
the system, and the $B$ individuals occupy the right half. In the
ensuing dynamics, the $A$ and $B$ individuals have identical
mobilities, growth rates, and intraspecific competition
parameters. However, the $A$ individuals are disadvantaged by having a
slightly higher death rate than the $B$'s. This is balanced by giving
the $A$'s an interspecific competitive advantage over the $B$'s. On
varying the strengths of these balancing forces it is possible to
obtain invasion of $A$'s from left to right or invasion of $B$'s from
right to left. We have chosen an example of the latter. It is seen
that mean field theory does an excellent job in predicting the long
time dynamics of the system.  In this figure the patch size is rather
large with a capacity of 100.

In Figure 7 we repeat the exact simulation as before but
simply reduce the patch size from 100 to 10. In this case the A
individuals are severely affected by discrete extinction events and
their density is in poor agreement with mean field
theory. Interestingly, the denser $B$ individuals are fairly well
described by mean field theory throughout the range.

We also studied an alternative balance of effects as follows. In
Figure 8 we show a situation in which the death rates for the
two species are the same, but now we reduce the amount of
intraspecific competition among the $B$ individuals. Again, the $A$'s
have an interspecific competitive advantage over the $B$'s. In this
case the $A$'s invade the $B$'s.  Here the patch size has the
intermediate value of 50 individuals. We see that mean field theory
performs relatively well.

On reducing the patch size for this particular scenario, from 50 to 10
(Figure 9) we see the failure of mean field theory (which
predicts invasion from left to right). The enhanced fluctuations in
the smaller patches lead to a quasi-dynamical balance in the
interfacial region between $A$'s and $B$'s. In this region the $A$'s
are beset by fluctuation-induced extinction events and this makes them
too weak to invade the $B$'s in the usual manner of a Fisher
wave. Instead, over longer scales than shown in the figure, the
density of $A$'s slowly permeates the $B$-rich region in a
``creeping'' motion.

\section{Conclusions}

There are many ways to formulate population dynamics. Popular
descriptions tend to be either deterministic (mean field) equations,
or individual based algorithms designed for implementation on a
computer. The extreme difference in these two approaches has led to
difficulties in directly comparing results. Disparities may be due to
fundamental deficiencies in one or both of the techniques, or else be
attributable to ``renormalization'' of various parameters. In this
paper we have attempted to bridge the gap between mean field models
and individual-based models.  We have described a very general
framework with which to formulate population dynamics using the
language of ``patches'' to create a concrete picture of the stochastic
process. The size of the patch is the central parameter. Mean field
theory is recovered on taking the patch size to infinity, while
discrete stochastic effects become prominent for small
patches containing a few individuals. Again, we emphasize that in our
usage ``mean field'' refers to the approximation in which
cross-correlations between stochastic variables is neglected, but
still allows for an explicitly spatial description.

From a biological perspective, a patch can be thought of as a (small)
spatial region within which interactions between individuals occur. It
is assumed that movement within this scale is not biologically
significant. In our spatial patch model, movement of an individual
between patches is biologically significant since that individual will
now have interactions with a new set of individuals in a neighboring
region. For systems in which interactions (not involving movement per
se) occur over larger scales, it will be necessary to include
additional inter-patch processes.

We have studied both non-spatial and spatial models. The non-spatial
case corresponds to a single patch containing a number of individuals
of both species. We have derived the corresponding mean field theory
and its first order corrections (i.e. Gaussian fluctuations about the
deterministic predictions). Generally, so long as the patch size is
above a critical value (which tends to be of the order of 100 in the
examples shown here), and the birth and death rates are such that a
sizable quasi-steady-state population is possible, then the mean-field
theory and its corrections give a satisfactory description of the
system. For smaller patches, or for situations in which there is a
non-negligible probability of extinction, it is crucial to account for
the discrete nature of the individuals. The population dynamics is
inherently stochastic and one must dispense with a deterministic
description. By tuning the patch size we have seen that the transition
from a mean-field like to a stochastic regime is rather sharp and
dependent on the existence of inter-specific interactions (in
this case, competition).

The same general picture holds for the spatially explicit models. We
have discussed two types of spatial patch models. In one, at each
spatial site there is a ``micro-patch'' which may hold at most one
individual. Movement and competition occurs between patches. In the
other, each lattice site is a patch of tunable size and competition occurs
inside the patch. Movement, of course, is still between patches. A
careful formulation shows that each model has the same spatial
mean-field limit. Of particular interest is the emergence of novel
non-linear diffusion terms, which are only present when two or more
species are competing for space. These terms are not written down in
the standard ``intuitively derived'' continuum equations of spatial
competition models. They are especially important in spatial regions
in which the density of one species is high, while the density of the
other is strongly spatially varying.  This would occur, for instance,
in a region of space containing a population boundary for one species
(due to some environmental barrier) but not for the other. We intend
to investigate such effects in more detail in a follow-up paper.

In our investigation of spatial mean field models, we have found them
to be more robust than in the non-spatial case. This is primarily due
to the weakening of local extinction via continual rescue effects from
neighboring patches. It is still the case however that as the patch
size is decreased, the quantitative precision of mean-field models
suffers, and with smaller patches still (we have in mind patches of size
10 or less) new stochastically driven qualitative features emerge. An
example of this was given (Figure 9) in which an invasion process (in
mean field theory) was halted due to stochastic weakening of the
leading edge of the invading population density.

In conclusion, we have presented a simple framework with which to
discuss fluctuation effects in population biology. This framework is
based on the use of patch models as concrete realizations of
stochastic processes.  The transition from mean field behavior to
fluctuation dominated stochastic dynamics is effected by reducing the
size of the patch. The critical patch size separating these two
regimes depends sensitively on the biological interactions
present. This has been an intensively theoretical work. In future work
we intend to apply the patch model to a variety of multi-species
population dynamics to address the importance of fluctuations and
validity of mean field theories in a quantitative and controlled
manner.

\acknowledgments 
We would like to acknowledge partial financial support from the National 
Science Foundation under grant number DEB-0328267 and the Jeffress 
Memorial Trust.

\appendix 
\section{Large $N$ analysis} 
\label{appA}

This appendix contains the details of the large $N$ analysis for the two 
species case which was described in section \ref{nonspatial} in the one 
species case,  

It is once again useful to write the master equation (\ref{two_master}) in 
the form
\begin{eqnarray}
\frac{dP(n,m,t)}{dt} & = & \left( {\cal E}_{x} - 1 \right)\, 
\left[ T(n-1,m|n,m)P(n,m,t) \right] \nonumber \\
& + & \left( {\cal E}^{-1}_{x} - 1 \right)\, 
\left[ T(n+1,m|n,m)P(n,m,t) \right] \nonumber \\
& + & \left( {\cal E}_{y} - 1 \right)\, 
\left[ T(n,m-1|n,m)P(n,m,t) \right] \nonumber \\
& + & \left( {\cal E}^{-1}_{y} - 1 \right)\, 
\left[ T(n,m+1|n,m)P(n,m,t) \right]
\label{twostep_master}
\end{eqnarray}
where the step operators $\cal E$ are defined by their actions on functions
of $n$ and $m$ by ${\cal E}^{\pm 1}_{x}f(n,m,t) = f(n \pm 1,m,t)$ and 
${\cal E}^{\pm 1}_{y}f(n,m,t) = f(n,m \pm 1,t)$. 

Writing $n = N\phi(t) + N^{1/2}\xi$ and $m = N\psi(t) + N^{1/2}\eta$, van 
Kampen's method yields the macroscopic equations
\begin{equation}
\frac{d\phi}{dt} = \alpha_{1,0}(\phi,\psi) \ \ \ \frac{d\psi}{dt} 
= \beta_{1,0}(\phi,\psi)
\label{two_macro}
\end{equation}
to leading order and the linear Fokker-Planck equation,
\begin{eqnarray}
\frac{\partial \Pi}{\partial t} & = & \left[ - \frac{\partial \alpha_{1,0}}
{\partial \phi} \right]\,\frac{\partial }{\partial \xi}\,\left( \xi\Pi \right) 
+ \left[ - \frac{\partial \alpha_{1,0}}{\partial \psi} \right]\,
\frac{\partial }{\partial \xi}\,\left( \eta\Pi \right) \nonumber \\ 
& + & \left[ - \frac{\partial \beta_{1,0}}
{\partial \phi} \right]\,\frac{\partial }{\partial \eta}\,\left( \xi\Pi \right)
+ \left[ - \frac{\partial \beta_{1,0}}{\partial \psi} \right]\,
\frac{\partial }{\partial \eta}\,\left( \eta\Pi \right) \nonumber \\
& + & \frac{1}{2} \alpha_{2,0}\,\frac{\partial^{2} \Pi}{\partial \xi^{2}}
+ \frac{1}{2} \beta_{2,0}\,\frac{\partial^{2} \Pi}{\partial \eta^{2}}
\label{two_FP}
\end{eqnarray}
to next order. This is a multivariate Fokker-Planck equation, but it is again
linear, and so its solution is a (multivariate) Gaussian. The $\alpha$ and
$\beta$ functions are given by
\begin{eqnarray}
\alpha_{1,0}(\phi,\psi) & = & 2 \tilde{b}_{1}\phi (1 - \phi - \psi ) 
- \left\{ \tilde{c}_{11}\phi^{2} + \tilde{d}_{1}\phi + 
2 \tilde{c}_{12}\phi\psi \right\} \nonumber \\
\beta_{1,0}(\phi,\psi) & = & 2 \tilde{b}_{2}\psi (1 - \phi - \psi ) 
- \left\{ \tilde{c}_{22}\psi^{2} + \tilde{d}_{2}\psi + 
2 \tilde{c}_{21}\phi\psi \right\} \nonumber \\
\alpha_{2,0}(\phi,\psi) & = & 2 \tilde{b}_{1}\phi (1 - \phi - \psi ) 
+ \left\{ \tilde{c}_{11}\phi^{2} + \tilde{d}_{1}\phi + 
2 \tilde{c}_{12}\phi\psi \right\} \nonumber \\
\beta_{2,0}(\phi,\psi) & = & 2 \tilde{b}_{2}\psi (1 - \phi - \psi ) 
+ \left\{ \tilde{c}_{22}\psi^{2} + \tilde{d}_{2}\psi + 
2 \tilde{c}_{21}\phi\psi \right\}
\label{al_and_be}
\end{eqnarray}

Since the solution to the Fokker-Planck equation is a Gaussian, we need once 
again only find the first two moments. They satisfy
\begin{eqnarray}
\frac{d }{dt}\,\langle \xi \rangle_{t} & = & \left[ +
\frac{\partial \alpha_{1,0}}{\partial \phi} \right]\,\langle \xi \rangle_{t}
+ \left[ + \frac{\partial \alpha_{1,0}}{\partial \psi} \right]\,
\langle \eta \rangle_{t} \label{av_x} \\
\frac{d }{dt}\,\langle \eta \rangle_{t} & = & \left[ + 
\frac{\partial \beta_{1,0}}{\partial \phi} \right]\,\langle \xi \rangle_{t}
+ \left[ + \frac{\partial \beta_{1,0}}{\partial \psi} \right]\,
\langle \eta \rangle_{t} \label{av_y} \\
\frac{d }{dt}\,\langle \xi^{2} \rangle_{t} & = & 2 \left[ + 
\frac{\partial \alpha_{1,0}}{\partial \phi} \right]\,
\langle \xi^{2} \rangle_{t}
+ 2 \left[ + \frac{\partial \alpha_{1,0}}{\partial \psi} \right]\,
\langle \xi \eta \rangle_{t} + \alpha_{2,0} \label{av_xx} \\
\frac{d }{dt}\,\langle \eta^{2} \rangle_{t} & = & 2 \left[ + 
\frac{\partial \beta_{1,0}}{\partial \phi} \right]\,\langle 
\xi \eta \rangle_{t}
+ 2 \left[ + \frac{\partial \beta_{1,0}}{\partial \psi} \right]\,
\langle \eta^{2} \rangle_{t} + \beta_{2,0} \label{av_yy} \\
\frac{d }{dt}\,\langle \xi \eta \rangle_{t} & = & \left[ + 
\frac{\partial \alpha_{1,0}}{\partial \phi} \right]\,
\langle \xi \eta \rangle_{t}
+ \left[ + \frac{\partial \alpha_{1,0}}{\partial \psi} \right]\,
\langle \eta^{2} \rangle_{t} \nonumber \\
& + & \left[ + \frac{\partial \beta_{1,0}}{\partial \phi} \right]\,
\langle \xi^{2} \rangle_{t} + \left[ + \frac{\partial \beta_{1,0}}
{\partial \psi} \right]\,\langle \xi \eta \rangle_{t} \label{av_xy}
\end{eqnarray}

We set the initial conditions on the macroscopic equations by asking that
\begin{equation}
\phi(0) = \frac{n_{0}}{N} \ \ \ \psi(0) = \frac{m_{0}}{N} 
\label{two_IC}
\end{equation}
This implies $\xi(0)=0$ and $\eta(0)=0$, and by successive differentiation of 
the macroscopic equations, that all derivatives of $\langle \xi \rangle_{t}$ 
and $\langle \eta \rangle_{t}$ at $t=0$ are also zero. We therefore take
\begin{equation}
\langle \xi \rangle_{t} = 0 \ \ \ \ \langle \eta \rangle_{t} = 0 
\label{x_and_y}
\end{equation}
for all $t$. Since the macroscopic equations with initial conditions 
(\ref{two_IC}) cannot be solved in closed form neither can the equations for
$\langle \xi^{2} \rangle_{t}, \langle \eta^{2} \rangle_{t}$ or 
$\langle \xi \eta \rangle_{t}$.

\section{Spatial models} 
\label{appB}

In this appendix we give details of the transition probabilities and the
master equations for the spatial models discussed in section \ref{spatial} 
of the main text. The results are frequently fairly straightforward 
generalizations of those found for the non-spatial model, however there are
some surprises in store: for example, the non-trivial spatial terms found
in the mean field theory of the two species model is only found by a careful
step-by-step derivation of the equation satisfied by 
$d\langle n_{i} \rangle/dt$.

We begin with the first version of the one-species model. The transition 
probabilities for the processes defined by (\ref{one_rates_i}) and 
(\ref{one_death}) are, by analogy with (\ref{one_transprobs}),
\begin{eqnarray}
T(\ldots n_{i}-1\ldots | \ldots n_{i}\ldots) & = & \frac{q_{1}c}{\Omega}\,
\frac{n_i}{N}\,\frac{(n_{i}-1)}{N-1} + \frac{(1-q_{1}-q_{2})d}{\Omega}\,
\frac{n_i}{N}, \nonumber \\
T(\ldots n_{i}+1\ldots | \ldots n_{i}\ldots) & = & \frac{2q_{1}b}{\Omega}\,
\frac{n_i}{N}\,\frac{(N-n_{i})}{N-1}\,. 
\label{one_transprobs_spat1.1}
\end{eqnarray}
The only change is the addition of the factor $\Omega^{-1}$, where $\Omega$ is
the number of sites in the lattice, which represents the arbitrary choice of 
the lattice site. In the case where the transition probabilities involve 
two neighboring patches, which is the process described by (\ref{one_mig}), the
corresponding quantities are
\begin{eqnarray}
T(\ldots n_{i}-1,n_{j}+1\ldots | \ldots n_{i},n_{j}\ldots) & = & 
\frac{q_{2}m}{z\Omega}\,\frac{n_i}{N}\,\frac{(N-n_{j})}{N} \nonumber \\
T(\ldots n_{i}+1,n_{j}-1\ldots | \ldots n_{i},n_{j}\ldots) & = & 
\frac{q_{2}m}{z\Omega}\,\frac{n_j}{N}\,\frac{(N-n_{i})}{N}\,, 
\label{one_transprobs_spat1.2}
\end{eqnarray}
where $z$ is the coordination number of the lattice (the number of nearest 
neighbors of any given site), and represents the choice of the nearest 
neighbor $j$, once $i$ has been chosen.
  
The master equation for this process therefore reads
\begin{eqnarray}
\frac{dP(\vec{n},t)}{dt} = \sum_{i}\,\sum_{j \in i} & \{ & 
T(\ldots n_{i},n_{j}\ldots | \ldots n_{i}+1,n_{j}-1 \ldots)
P(\ldots n_{i}+1,n_{j}-1\ldots,t)\nonumber \\
& + & T(\ldots n_{i},n_{j}\ldots | \ldots n_{i}-1,n_{j}+1 
\ldots)P(\ldots n_{i}-1,n_{j}+1\ldots,t) \ \ \} \nonumber \\ \nonumber \\
+ \sum_{i} & \{ & T(\ldots n_{i}\ldots | \ldots n_{i}+1\ldots)
P(\ldots n_{i}+1\ldots,t)\nonumber \\ 
& + & T(\ldots n_{i}\ldots | \ldots n_{i}-1\ldots)
P(\ldots n_{i}-1\ldots,t) \ \ \} \nonumber \\ \nonumber \\ 
- \sum_{i}\,\sum_{j \in i} & \{ & 
T(\ldots n_{i}-1,n_{j}+1\ldots | \ldots n_{i},n_{j}\ldots)
P(\ldots n_{i},n_{j}\ldots,t)\nonumber \\ 
& + & T(\ldots n_{i}+1,n_{j}-1\ldots | \ldots n_{i},n_{j}\ldots)
P(\ldots n_{i},n_{j}\ldots,t) \ \ \} \nonumber \\ \nonumber \\ 
- \sum_{i} & \{ & T(\ldots n_{i}-1\ldots | \ldots n_{i}\ldots)
P(\ldots n_{i}\ldots,t)\nonumber \\
& + & T(\ldots n_{i}+1\ldots | \ldots n_{i}\ldots)
P(\ldots n_{i}\ldots,t) \ \ \} . 
\label{one_master_spat}
\end{eqnarray}
Although this looks rather complicated, it is a straightforward generalization 
of (\ref{one_master}). In an effort to keep it as simple as possible, only the 
number of individuals at sites where changes occur ($i$ or $j$) have been 
explicitly shown on the right-hand side of the equation. The notation 
$j \in i$ denotes a sum over all sites $j$ which are nearest-neighbors of $i$. 
On the left-hand side of the equation, $\vec{n}$ denotes the number of 
individuals in the set of all patches: 
$\vec{n} \equiv (n_{1},\ldots,n_{i},\ldots,n_{j},\ldots)$.

To obtain the rate equation, we substitute (\ref{one_master_spat}) into
\begin{equation}
\frac{d\langle n_{k} \rangle}{dt} = \sum_{ \{ \vec{n} \} } n_{k}
\frac{dP(\vec{n},t)}{dt}\,.
\label{one_rate_spat}
\end{equation}
Defining new quantities
\begin{equation}
\tilde{c} = \frac{q_{1} c}{N-1}, \ \ \ \tilde{b} = \frac{q_{1} b}{N-1}, 
\ \ \  \tilde{d} = \frac{(1-q_{1} - q_{2})d}{N}, \ \ \ \tilde{m} = 
\frac{q_{2} m}{N}\,,
\label{one_tilde_spat}
\end{equation}
as in (\ref{one_tilde}), and introducing a rescaled time $\tau = t/\Omega$, 
the following equation is found:
\begin{equation}
\frac{d\ }{d\tau} \left\langle \frac{n_{i}}{N} \right\rangle= 
\tilde{m}\,\Delta \left\langle \frac{n_{i}}{N} \right\rangle 
- \tilde{c} \left\{ \left\langle \frac{n_{i}^{2}}{N^2} \right\rangle 
- \frac{1}{N}\,\left\langle \frac{n_{i}}{N} \right\rangle \right\} 
+ 2\tilde{b} \left\{ \left\langle \frac{n_{i}}{N} \right\rangle
- \left\langle \frac{n_{i}^{2}}{N^2} \right\rangle \right\} 
- \tilde{d} \left\langle \frac{n_{i}}{N} \right\rangle \,,
\label{one_noverN_spat1}
\end{equation}
where we have used the explicit forms (\ref{one_transprobs_spat1.1}) and 
(\ref{one_transprobs_spat1.2}). The symbol $\Delta$ denotes the lattice 
Laplacian (with unit lattice spacing):
\begin{equation}
\Delta f_{i} \equiv \frac{2}{z}\,\sum_{j \in i} 
\left( f_{j} - f_{i} \right) \,.
\label{lattice_lap}
\end{equation}
The corresponding population-level description can be obtained from 
(\ref{one_noverN_spat1}) by letting $N \to \infty$ which eliminates the term 
of order $N^{-1}$ and allows us to replace $\langle n_{i}^{2} \rangle$
by $\langle n_{i} \rangle^{2}$, as in section \ref{basic}. This leads to
an equation for $\phi_{i} \equiv \langle n_{i} \rangle/N$ which is given by
\begin{equation}
\frac{d\phi_{i}}{d\tau} = \tilde{m} \Delta \phi_{i} 
- \tilde{c} \phi^{2}_{i} + 2\tilde{b} \phi_{i} \left(1 - \phi_{i} \right)
- \tilde{d} \phi_{i}\,.
\label{one_phieqn_spat}
\end{equation}
The final step that has to be taken in order to make contact with the 
equations used in the traditional approach, is to move from the lattice to 
the continuum. To do this we need to introduce a lattice spacing of $\epsilon$ 
and take it to zero so that
\begin{equation}
\lim_{\epsilon \to 0}\,\frac{2}{z}\,\sum_{j \in i} 
\frac{( \phi_{j} - \phi_{i} )}
{\epsilon^{2}} \longrightarrow \nabla^{2}\phi ({\bf x}) \,,
\label{continuum}
\end{equation}
where the lattice site $i$ is now replaced by the position vector ${\bf x}$.
In addition, the migration parameter has to be redefined, in order to absorb 
a factor of $\epsilon^{2}$. The resulting equation is (\ref{one_phi_eqn_spat}),
given in the main text.

The derivation of the population-level description for the second version of
the one-species model follows similar lines. The particular differences 
between this version and the one discussed above are described in the main 
text, and specifically by the equations (\ref{one_bcm}) and 
(\ref{one_ver2_comp}). The transition probabilities for this second version
are
\begin{eqnarray}
T(\ldots n_{i}-1,n_{j}\ldots | \ldots n_{i},n_{j}\ldots) 
&=& \frac{\mu c}{zN}\,n_{i}n_{j} \nonumber \\
T(\ldots n_{i}+1,n_{j}\ldots | \ldots n_{i},n_{j}\ldots)
&=& \frac{\mu b}{zN}\,(1-n_{i})n_{j} \nonumber \\
T(\ldots n_{i}+1,n_{j}-1\ldots | \ldots n_{i},n_{j}\ldots)
&=& \frac{\mu m}{zN}\,(1-n_{i})n_{j} \,,
\label{one_transprobs_spat2.1}
\end{eqnarray}
with similar equations with $i$ and $j$ interchanged, and 
\begin{equation} 
T(\ldots n_{i}-1\ldots | \ldots n_{i}\ldots)
= \frac{(1-\mu) d}{N}\,n_{i} \,.
\label{one_transprobs_spat2.2}
\end{equation}
Note that the transition probabilities in (\ref{one_transprobs_spat2.1}) are
zero unless $n_{i}$ and $n_{j}$ are both equal to 1 (competition) or
$n_{i}=0$ and $n_{j}=1$ (birth and migration), as required. The factors $zN$
and $N$ account for the choices of sites $i$ and $j$ and replace $z\Omega$
and $\Omega$ respectively in the first version.

The master equation resembles (\ref{one_master_spat}), except that the 
single site processes are now restricted to the death process, and the 
two-site processes are more extensive, involving birth, competition, and 
migration. Defining
\begin{equation}
\hat{c} = \frac{\mu c}{N}, \ \ \ \hat{b} = \frac{\mu b}{N}, 
\ \ \  \hat{d} = \frac{(1-\mu)d}{N}, \ \ \ \hat{m} = \frac{\mu m}{N}
\ \ \ \tau = \frac{t}{N}\,,
\label{one_hat_spat}
\end{equation} 
we find using (\ref{one_rate_spat}) and the decoupling approximation
$\langle n_{i} n_{j} \rangle = \langle n_{i} \rangle \langle n_{j} \rangle$,
\begin{equation}
\frac{d\ }{d\tau} \langle n_{i} \rangle = \hat{m}\,\Delta 
\langle n_{i} \rangle - \hat{c} \langle n_{i} \rangle \left\{ \frac{1}{z} 
\sum_{j \in i} \langle n_{j} \rangle \right\} +
2 \hat{b} \left( 1 - \langle n_{i} \rangle \right) \left\{ \frac{1}{z}
\sum_{j \in i} \langle n_{j} \rangle \right\} - \hat{d} \langle n_{i} \rangle
\,. 
\label{one_<n>_spat2}
\end{equation}
Denoting $\langle n_{i} \rangle$ as $\phi_{i}$, the terms in curly brackets
become $\phi ({\bf x},t)$ in the continuum limit, and so once again we 
recover equation (\ref{one_phi_eqn_spat}), given in the main text.

The description of the IBMs when two species are present parallels that for 
one species. This similarity also holds for the initial stages of the 
derivation of the population-based equations, and so our description will be 
brief for both of these aspects. 

For the first version of the model, the transition probabilities for birth,
competition and death processes are generalizations of (\ref{two_transprobs})
(the modifications are exactly the same as those made on (\ref{one_transprobs})
to give (\ref{one_transprobs_spat1.1})). Those for migration of $A$'s are 
(\ref{one_transprobs_spat1.2}), but with $m$ replaced by $m_{1}$ and 
$N - n_{j}$ replaced by $N - n_{j} - m_{j}$ (or $N - n_{i}$ replaced by 
$N - n_{i} - m_{i}$). For migration of $B$'s, they have the same form, but 
with $m_{1}$ replaced by $m_{2}$ and with the substitutions 
$n_{i} \leftrightarrow m_{i}$ and $n_{j} \leftrightarrow m_{j}$. The 
master equation for $P(\vec{n}, \vec{m}, t)$ is as before, but now including 
the greater number of allowed processes. There are two rate equations, found 
by substituting the master equations into
\begin{equation}
\frac{d\langle n_{k} \rangle}{dt} = \sum_{ \{ \vec{n} \} } 
\sum_{ \{ \vec{m} \} } n_{k}\frac{dP(\vec{n}, \vec{m}, t)}{dt} \ \ {\rm and} 
\ \ \frac{d\langle m_{k} \rangle}{dt} = \sum_{ \{ \vec{n} \} } 
\sum_{ \{ \vec{m} \} } m_{k}\frac{dP(\vec{n}, \vec{m}, t)}{dt}\,.
\label{two_rate_spat}
\end{equation}
Defining new quantities
\begin{equation}
\tilde{c}_{\alpha \beta} = \frac{q_{1} c_{\alpha \beta}}{N-1}, \ \ \ 
\tilde{b}_{\alpha} = \frac{q_{1} b_{\alpha}}{N-1}, \ \ \  \tilde{d}_{\alpha} = 
\frac{(1-q_{1} - q_{2})d_{\alpha}}{N}, \ \ \ \tilde{m}_{\alpha} = 
\frac{q_{2} m_{\alpha}}{N}, \ \ \  \tau = \frac{t}{\Omega}\,,
\label{two_tilde_spat}
\end{equation}
as in (\ref{one_tilde_spat}), we now let $N \to \infty$ and replace averages 
of products by products of averages to obtain the equations
\begin{equation}
\frac{d\phi_{i}}{d\tau} = \tilde{m}_{1} \Delta \phi_{i} 
+ \frac{2\tilde{m}_{1}}{z} \sum_{j \in i} \left( \phi_{i} \psi_{j} - 
\phi_{j} \psi_{i} \right) - \tilde{c}_{11} \phi^{2}_{i} - 
2\tilde{c}_{12} \phi_{i} \psi_{i} + 2\tilde{b}_{1} \phi_{i} \left(1 - \phi_{i} 
- \psi_{i} \right) - \tilde{d}_{1} \phi_{i}
\label{two_phieqn_spat}
\end{equation}
and
\begin{equation}
\frac{d\psi_{i}}{d\tau} = \tilde{m}_{2} \Delta \psi_{i} 
+ \frac{2\tilde{m}_{2}}{z} \sum_{j \in i} \left( \psi_{i} \phi_{j} - 
\psi_{j} \phi_{i} \right) - \tilde{c}_{22} \psi^{2}_{i} - 
2\tilde{c}_{21} \psi_{i} \phi_{i} + 2\tilde{b}_{2} \psi_{i} \left(1 - \phi_{i} 
- \psi_{i} \right) - \tilde{d}_{2} \psi_{i}\,.
\label{two_psieqn_spat}
\end{equation}
Here $\phi_{i} \equiv \langle n_{i} \rangle/N$ and 
$\psi_{i} \equiv \langle m_{i} \rangle/N$. Writing 
$\phi_{i} \psi_{j} - \phi_{j} \psi_{i}$ as
$\phi_{i} (\psi_{j} - \psi_{i}) - \psi_{i} (\phi_{j} - \phi_{i})$ we obtain
equations (\ref{two_phi_eqn_spat}) and (\ref{two_psi_eqn_spat}) in the 
continuum limit.

For the second version of the two species model, the transition probabilities
are generalizations of the one-species forms given by 
(\ref{one_transprobs_spat2.1}) and (\ref{one_transprobs_spat2.2}). 
Specifically, for the competition process the term $c n_{i} n_{j}$ becomes 
$c_{11} n_{i} n_{j}$ and, in addition, there are transition probabilities 
which are proportional to $c_{12} n_{i} m_{j}, c_{21} m_{i} n_{j}$ and
$c_{22} m_{i} m_{j}$. For the birth process, the factor $(1-n_{i})$ is 
replaced by $(1-n_{i}-m_{i})$, and $b n_{j}$ by $b_{1} n_{j}$ or 
$b_{2} m_{j}$. The same holds for migration, but with $b, b_{1}$ and $b_{2}$ 
replaced by $m, m_{1}$ and $m_{2}$ respectively. Finally, for the death 
process $d n_{i}$ is replaced by $d_{1} n_{i}$ or $d_{2} m_{i}$. The master 
equation is straightforward, but tedious, to write down.

Defining new quantities
\begin{equation}
\hat{c}_{\alpha \beta} = \frac{\mu c_{\alpha \beta}}{N}, \ \ \ 
\hat{b}_{\alpha} = \frac{\mu b_{\alpha}}{N}, \ \ \  \hat{d}_{\alpha} = 
\frac{(1-\mu) d_{\alpha}}{N}, \ \ \ \hat{m}_{\alpha} = 
\frac{\mu m_{\alpha}}{N}, \ \ \ \tau = \frac{t}{N}\,,
\label{two_hat_spat}
\end{equation}
we find using the decoupling approximation --- in which averages of products 
of any two of the variables $\{ \vec{n}, \vec{m} \}$ are replaced by the 
products of their averages --- that
\begin{eqnarray}
\frac{d\ }{d\tau} \langle n_{i} \rangle &=& \hat{m}_{1} \Delta 
\langle n_{i} \rangle + \frac{2 \hat{m}_{1}}{z} \left[ \langle n_{i} \rangle
\left\{ \frac{1}{z} \sum_{j \in i} \langle m_{j} \rangle \right\} - 
\langle m_{i} \rangle \left\{ \frac{1}{z} \sum_{j \in i} 
\langle n_{j} \rangle \right\} \right]  
- \hat{c}_{11} \langle n_{i} \rangle \left\{ \frac{1}{z} 
\sum_{j \in i} \langle n_{j} \rangle \right\} \nonumber \\
&-& 2 \hat{c}_{12} \langle n_{i} \rangle \left\{ \frac{1}{z} 
\sum_{j \in i} \langle m_{j} \rangle \right\} +
2 \hat{b}_{1} \left( 1 - \langle n_{i} \rangle - \langle m_{i} \rangle \right) 
\left\{ \frac{1}{z} \sum_{j \in i} \langle n_{j} \rangle \right\} - 
\hat{d}_{1} \langle n_{i} \rangle\,, 
\label{two_<n>_spat2}
\end{eqnarray}
and
\begin{eqnarray}
\frac{d\ }{d\tau} \langle m_{i} \rangle &=& \hat{m}_{2} \Delta 
\langle m_{i} \rangle + \frac{2 \hat{m}_{2}}{z} \left[ \langle m_{i} \rangle
\left\{ \frac{1}{z} \sum_{j \in i} \langle n_{j} \rangle \right\} - 
\langle n_{i} \rangle \left\{ \frac{1}{z} \sum_{j \in i} 
\langle m_{j} \rangle \right\} \right]
- \hat{c}_{22} \langle m_{i} \rangle \left\{ \frac{1}{z} 
\sum_{j \in i} \langle m_{j} \rangle \right\} \nonumber \\
&-& 2 \hat{c}_{21} \langle m_{i} \rangle \left\{ \frac{1}{z} 
\sum_{j \in i} \langle n_{j} \rangle \right\} +
2 \hat{b}_{2} \left( 1 - \langle n_{i} \rangle - \langle m_{i} \rangle \right) 
\left\{ \frac{1}{z} \sum_{j \in i} \langle m_{j} \rangle \right\} - 
\hat{d}_{2} \langle m_{i} \rangle\,. 
\label{two_<m>_spat2}
\end{eqnarray}
Defining $\langle n_{i} \rangle$ and $\langle m_{i} \rangle$ as $\phi_{i}$ 
and $\psi_{i}$ respectively, we recover equations (\ref{two_phi_eqn_spat}) and
(\ref{two_psi_eqn_spat}) in the continuum limit, up to slightly different 
definitions of the birth, competition, migration, and death rates.

\newpage

\noindent
{\bf \underline{Tables}}

\vspace{0.25in}

\begin{center}
\begin{tabular}{|c||c|c|c|c|c|c|}\hline
{Fig.} & {samples} & {$N$} & {$\mu $} & 
{$b$} & {$d$} & {$c$} \\ \hline
  1 & 10 000 & 100 & 0.5 & 0.5 & 0.5 & 0.5 \\ \hline
  2 & 50 000 & 10  & 0.5 & 0.5 & 0.5 & 0.5 \\ \hline
\end{tabular}
\end{center}
\begin{center}
\underline{Table 1}: Parameter values for Figures 1 and 2.
\end{center}

\vspace{0.5in}

\begin{center}
\begin{tabular}{|c||c|c|c|c|c|c|c|c|c|c|c|}\hline
{Fig.} & {samples} & {$N$} & {$\mu $} & 
{$b_{1}$} & {$b_{2}$} & {$d_{1}$} & {$d_{2}$} & {$c_{11}$} & {$c_{22}$} 
& {$c_{12}$} & {$c_{21}$} \\ \hline
  3 & 1000 & 400 & 0.5 & 0.5 & 0.5 & 0.5 & 0.5 & 0.5 & 0.5 & 0.0 & 0.0 
\\ \hline
  4 & 1000 & 200 & 0.5 & 0.5 & 0.5 & 0.5 & 0.5 & 0.5 & 0.5 & 0.0 & 0.0
\\ \hline
  5 & 1000 & 400 & 0.5 & 0.5 & 0.5 & 0.5 & 0.5 & 0.5 & 0.5 & 0.0 & 0.1
\\ \hline
\end{tabular}
\end{center}
\begin{center}
\underline{Table 2}: Parameter values for Figures 3-5.
\end{center}

\vspace{0.5in}

\begin{center}
\begin{tabular}{|c||c|c|c|c|c|c|c|c|c|c|c|c|c|c|c|}\hline
{Fig.} & {samples} & {$L$} & {$N$} & {$q_{1}$} & {$q_{2}$} & 
{$b_{1}$} & {$b_{2}$} & {$d_{1}$} & {$d_{2}$} & {$c_{11}$} & {$c_{22}$} 
& {$c_{12}$} & {$c_{21}$} & {$m_{1}$} & {$m_{2}$} \\ \hline
  6 & 250  & 100 & 100 & 1/3 & 1/3 & 0.5 & 0.5 & 0.7 & 0.5 & 0.5 & 0.5 & 0.0 
& 0.5 & 1.0 & 1.0 \\ \hline
  7 & 1000 & 100 & 10 &  1/3 & 1/3 & 0.5 & 0.5 & 0.7 & 0.5 & 0.5 & 0.5 & 0.0
& 0.5 & 1.0 & 1.0 \\ \hline
  8 & 500  & 100 & 50 &  1/3 & 1/3 & 0.5 & 0.5 & 0.5 & 0.5 & 0.5 & 0.2 & 0.0
& 0.5 & 1.0 & 1.0 \\ \hline
  9 & 500  & 100 & 10 &  1/3 & 1/3 & 0.5 & 0.5 & 0.5 & 0.5 & 0.5 & 0.2 & 0.0
& 0.5 & 1.0 & 1.0 \\ \hline
\end{tabular}
\end{center}
\begin{center}
\underline{Table 3}: Parameter values for Figures 6-9.
\end{center}

\newpage

\noindent
{\bf \underline{Figure Captions}}

\vspace{0.25in}

\noindent
\underline{Figure 1} A comparison of theory to simulation
for a single patch containing a single species. The upper panel
shows the time evolution of the population density 
$\phi(t)=\langle n \rangle/N$, where $n$ is the number of individuals 
and $N$ is the size of the patch. The lower panel shows the time evolution 
of the variance,
$v(t) = \left( \langle n^{2} \rangle - \langle n \rangle^{2} \right)/N^{2}$,
of the population. The subscript $A$ refers to the fact that the individuals 
belong to species $A$. In this case the patch size has the relatively large 
value of 100 and we see that theory and simulation are in good agreement. See
Table 1 for specific parameter values used in Figures 1 and 2.

\bigskip

\noindent
\underline{Figure 2} The same as Figure 1 but with
the patch size reduced to 10. In this case the mean population density
falls to zero due to fluctuation induced extinctions. The true
variance first exceeds the large $N$ prediction and then falls steeply
below due to the growing number of realizations which have become
extinct.

\bigskip

\noindent
\underline{Figure 3} Comparison of theory to simulation
for a patch containing two species, $A$ and $B$. Initially each species
has a density of one quarter of the total patch capacity. The upper
panel shows the time evolution of the densities of $A$ and $B$ ($\phi(t)$
and $\psi(t)$, respectively), while the lower panel shows the time evolution 
of the variance of $A$ compared with the large-$N$ theory. In this case the 
$A$ and $B$ individuals have identical birth, death, and competition 
parameters, and the interspecific competition is set to zero. The patch 
size has the large value of 400. See Table 2 for specific parameter values 
used in Figures 3-5.

\bigskip

\noindent
\underline{Figure 4} The same as Figure 3 but with a patch
size of 200. Note that although the mean field theory still works
fairly well, the large-$N$ prediction for the variance is very poor.
Even for such a large patch, the fluctuations are increasing steadily
with time.

\bigskip

\noindent
\underline{Figure 5} A similar scenario to Figure 3, with
the addition of weak interspecific interactions (one fifth of the
strength of the intraspecific interactions, with $A$ out-competing $B$).
Here, the patch size is 400, and the densities are fairly well approximated
by mean field theory, although we see a slow decline in the $B$ population. 
However, the fluctuations are again increasing with time and are not 
well-described by the large-$N$ theory for large times.

\bigskip

\noindent
\underline{Figure 6} Comparison of mean field theory (smooth lines) 
and simulation (erratic lines) for two species $A$ and $B$ in
a spatial setting in which initially the $A$ individuals occupy the
left half of the system and the $B$ individuals the right half.
The upper (lower) panel shows early (late) times. Here, the patch
size has the relatively large value of 100. $A$ out-competes $B$ (meaning
that $c_{21} > c_{12}$) but has a higher death rate, and so is invaded by $B$.
See Table 3 for specific parameter values used in Figures 6-9.

\bigskip

\noindent
\underline{Figure 7} The same as Figure 6 but here the 
patch size is reduced from 100 to the relatively small value of 10.
Note that mean field theory is still in fairly good agreement with
the high density $B$ population, but shows significant deviation for
the stochastically weakened $A$ population.

\bigskip

\noindent
\underline{Figure 8} A similar scenario to Figure 6, but
now $A$ and $B$ have identical death rates and yet $B$ has less intraspecific
competition than $A$. In this example $A$ invades $B$. The patch size here
has the relatively large value of 50.

\bigskip

\noindent
\underline{Figure 9} The same as Figure 8 except that the
patch size is reduced from 50 to 10. Note that the invasion of $A$ into
$B$ is severely slowed due to the stochastic weakening of $A$.

\newpage

\begin{figure}
\includegraphics[width=5.5in]{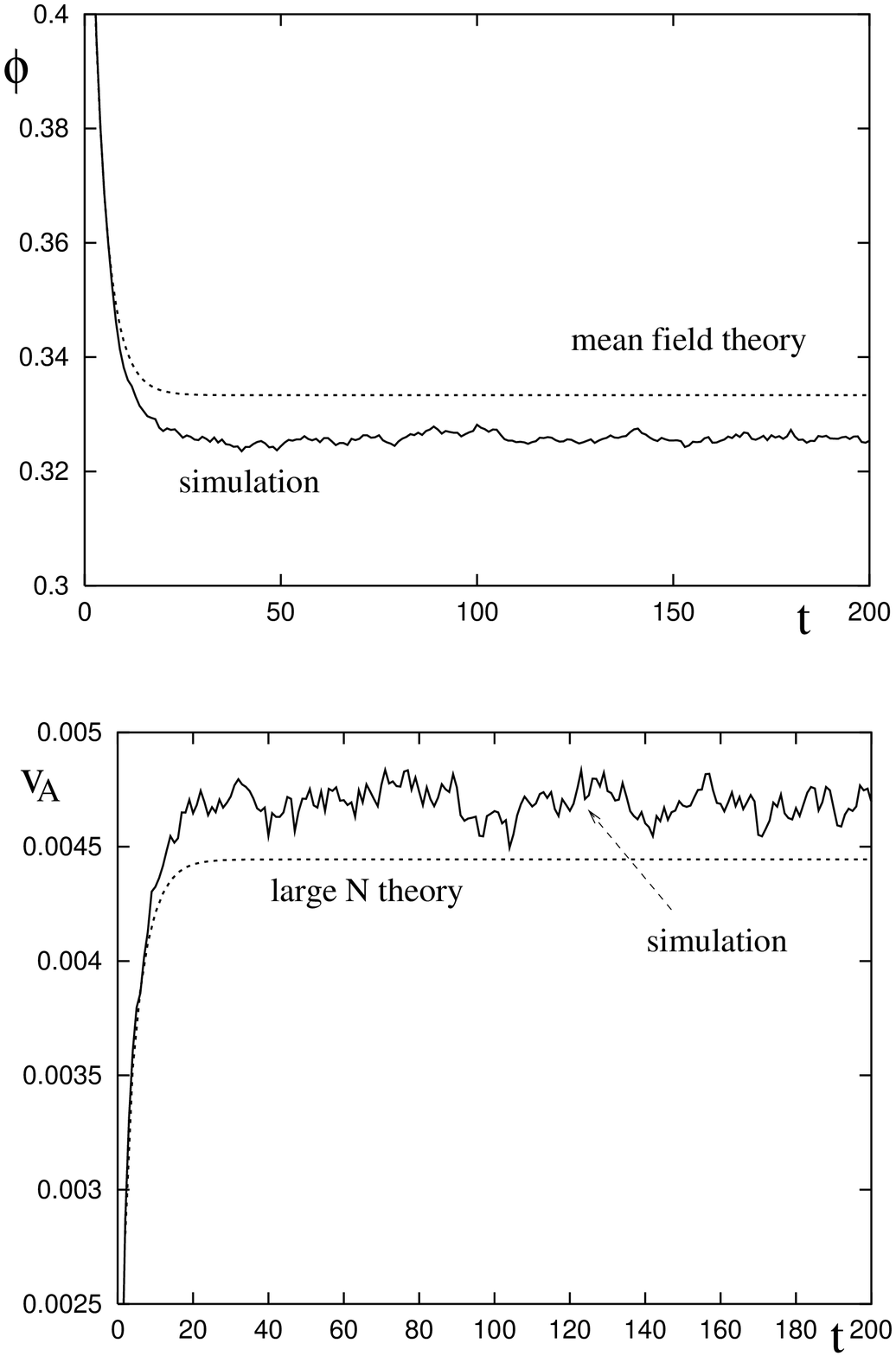} 
\caption{}
\end{figure}

\begin{figure}
\includegraphics[width=5.5in]{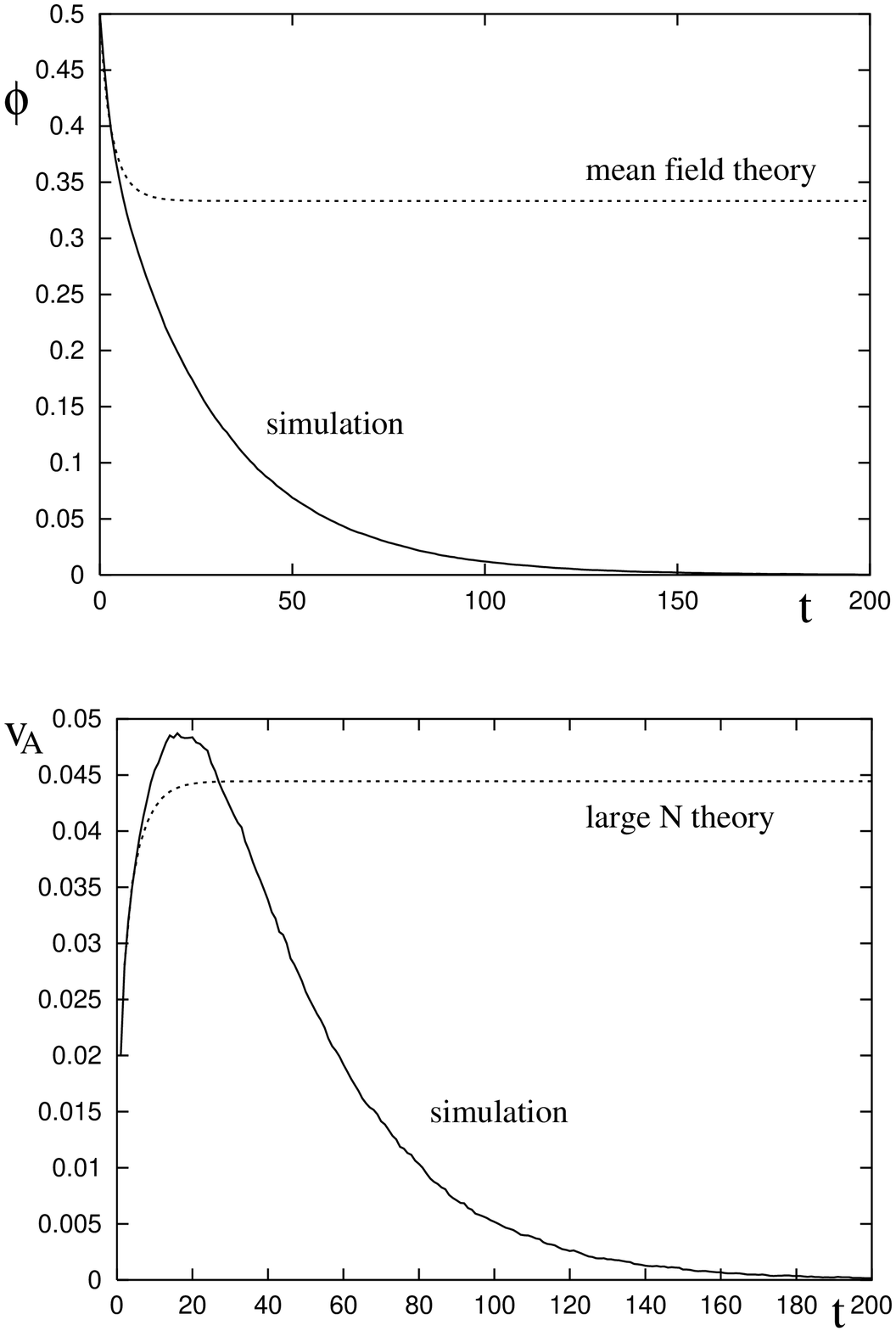} 
\caption{}
\end{figure}

\begin{figure}
\includegraphics[width=5.5in]{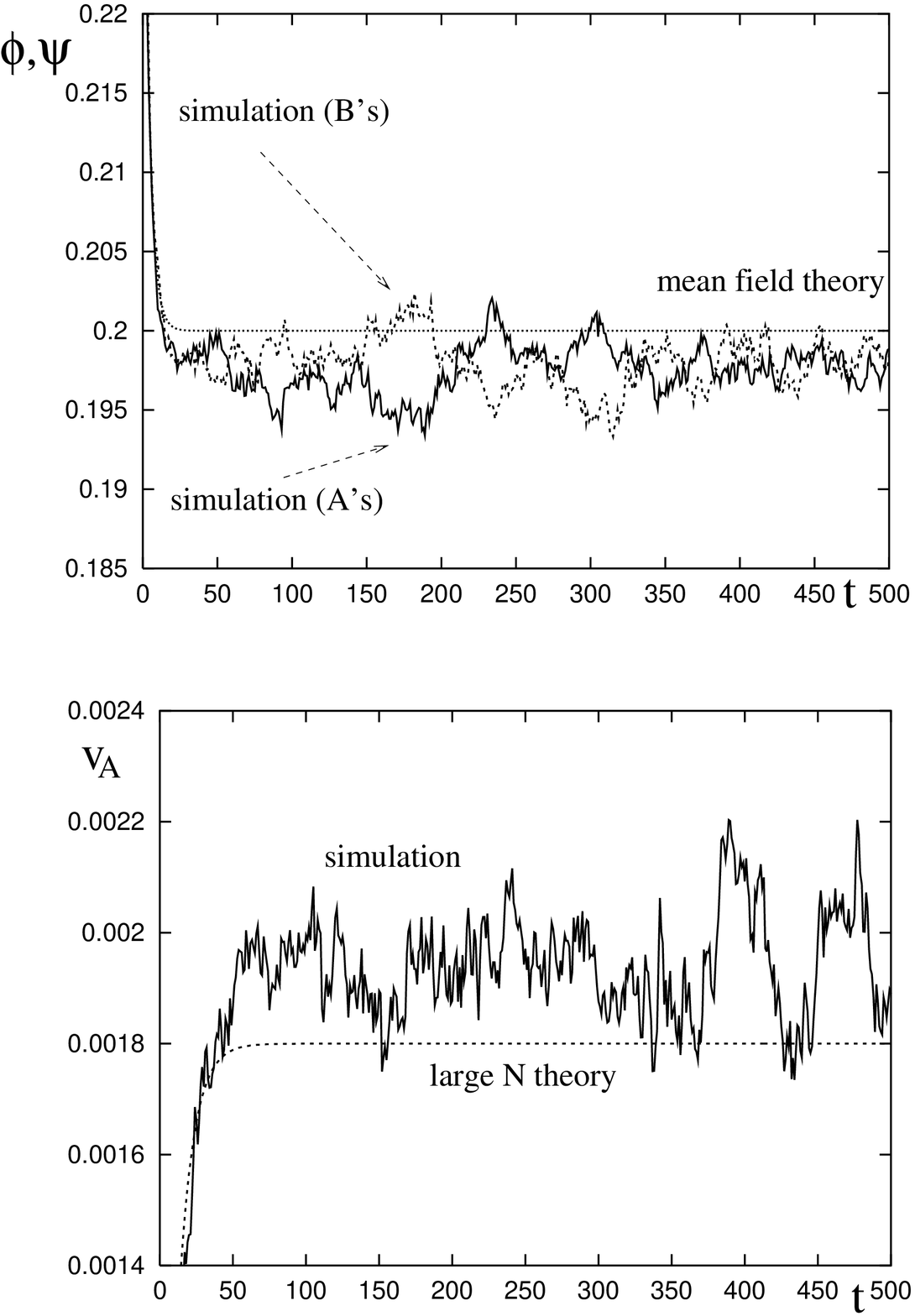} 
\caption{}
\end{figure}

\begin{figure}
\includegraphics[width=5.5in]{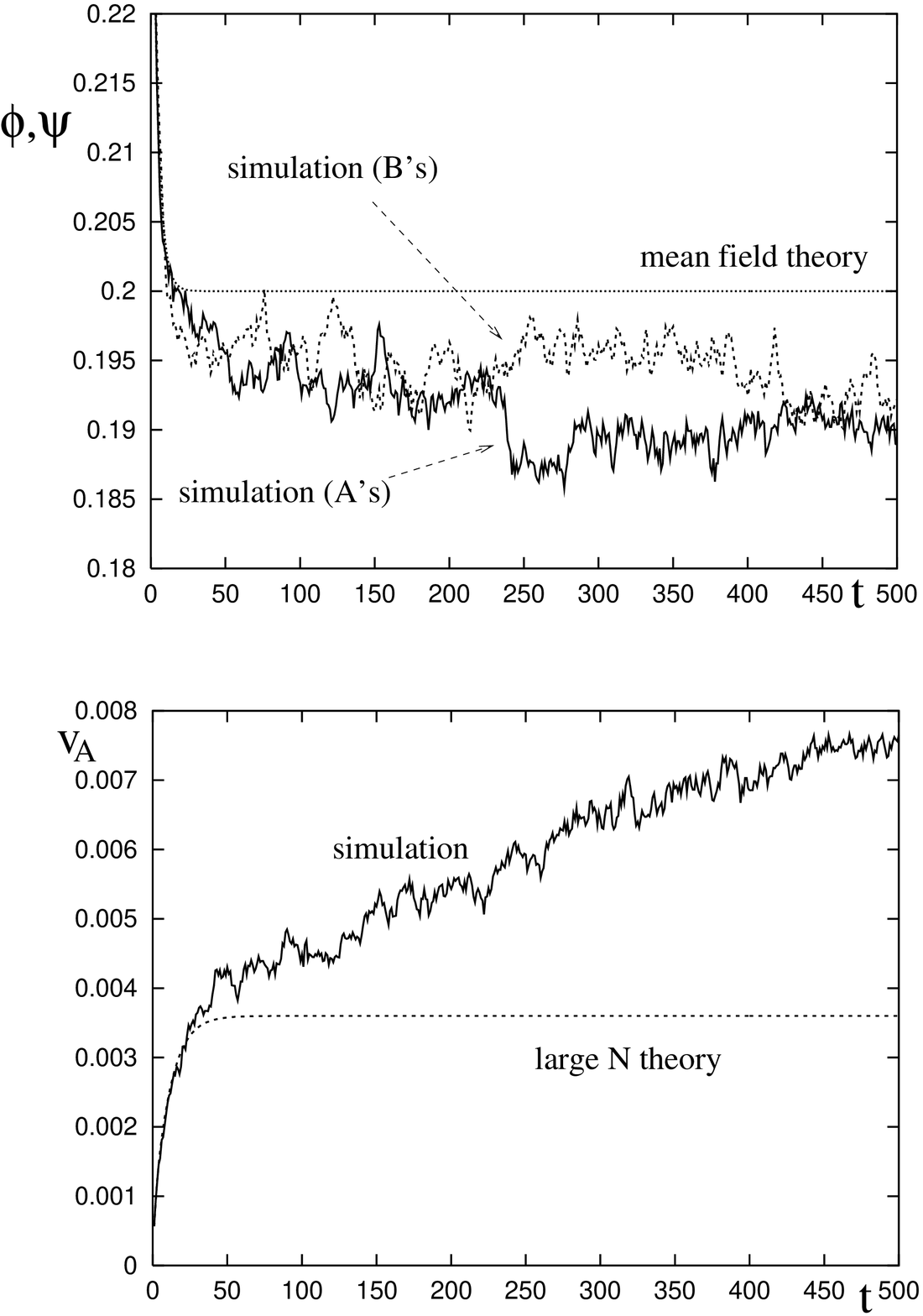} 
\caption{}
\end{figure}

\begin{figure}
\includegraphics[width=5.5in]{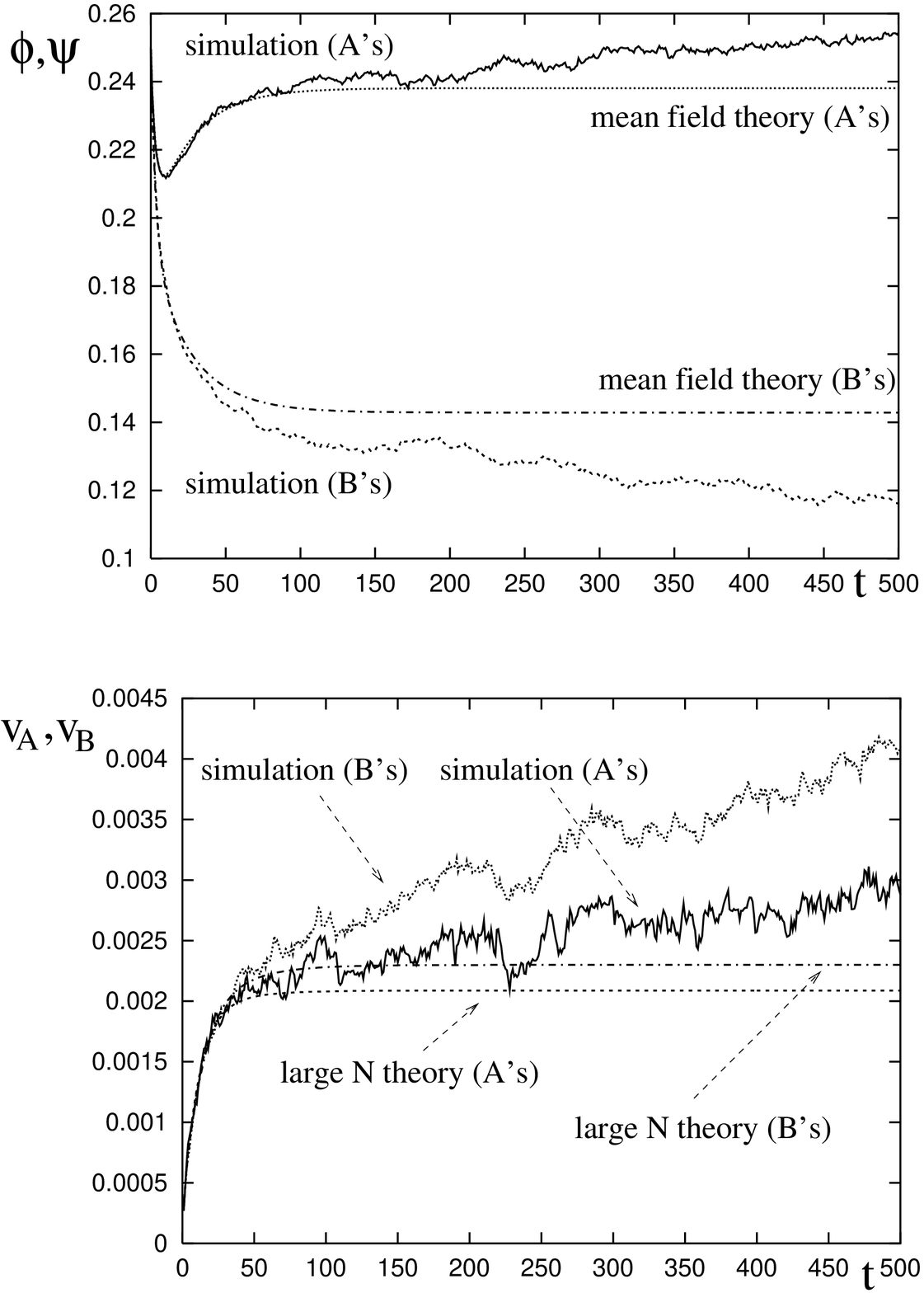} 
\caption{}
\end{figure}

\begin{figure}
\includegraphics[width=5.5in]{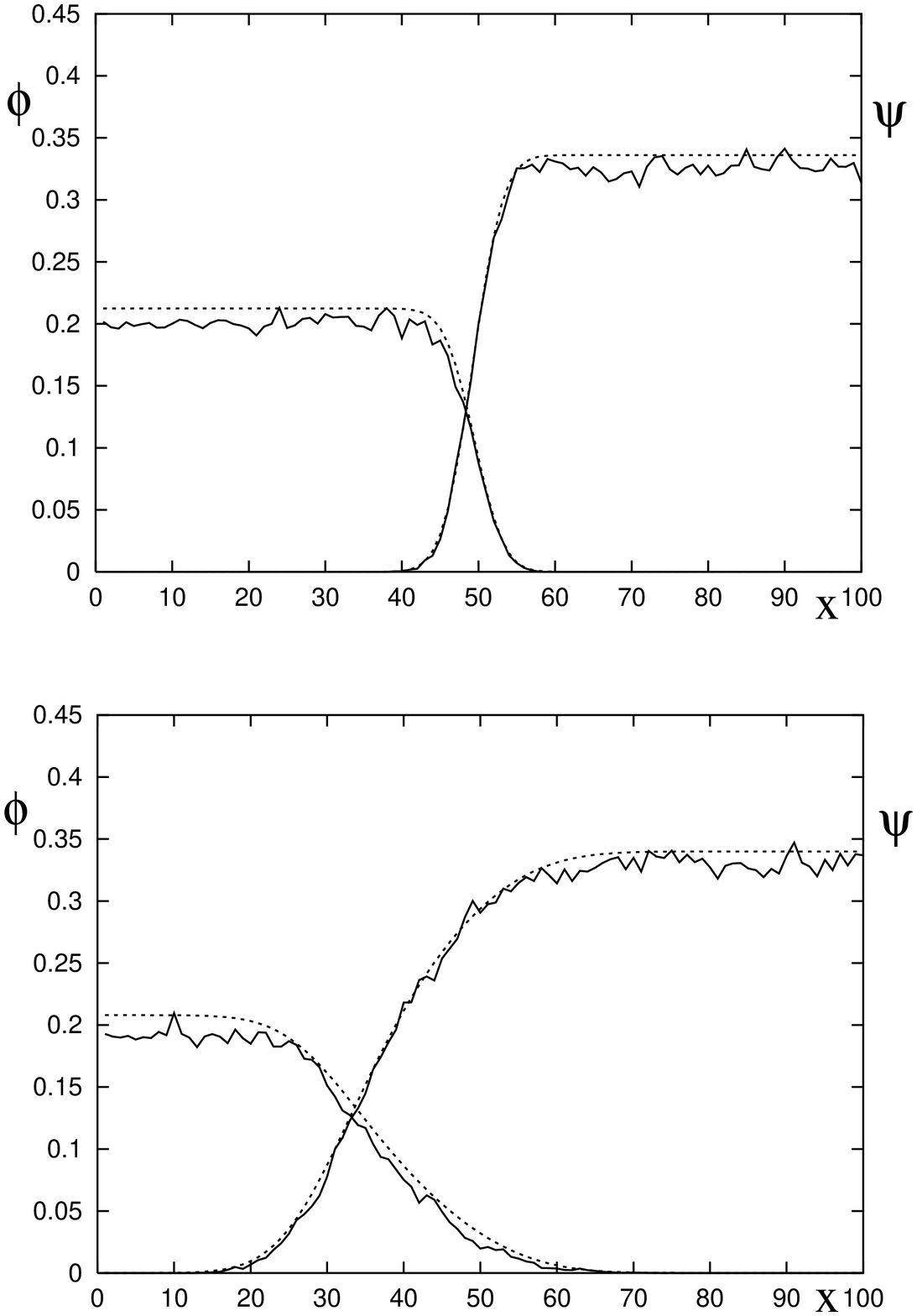} 
\caption{}
\end{figure}

\begin{figure}
\includegraphics[width=5.5in]{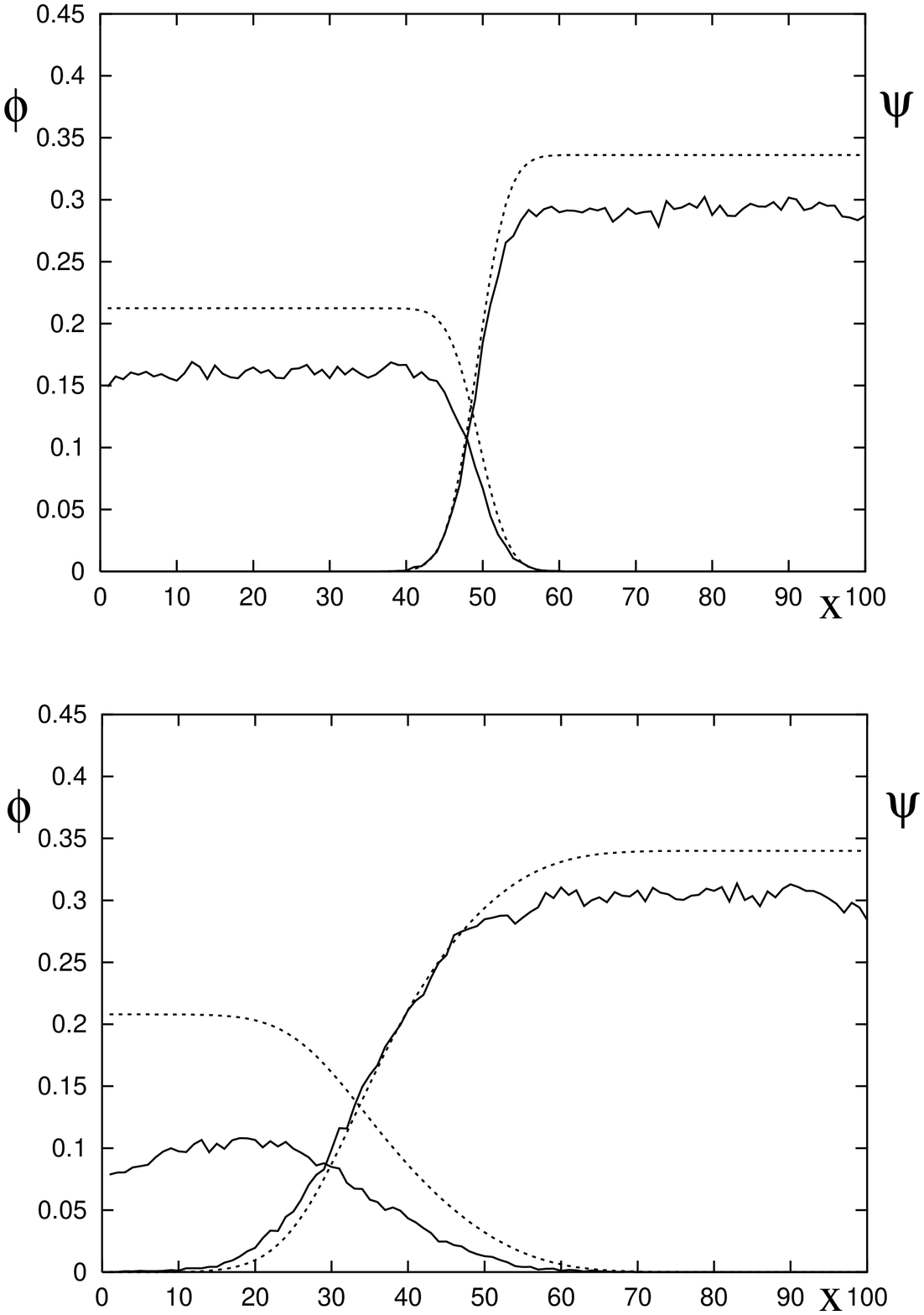} 
\caption{}
\end{figure}

\begin{figure}
\includegraphics[width=5.5in]{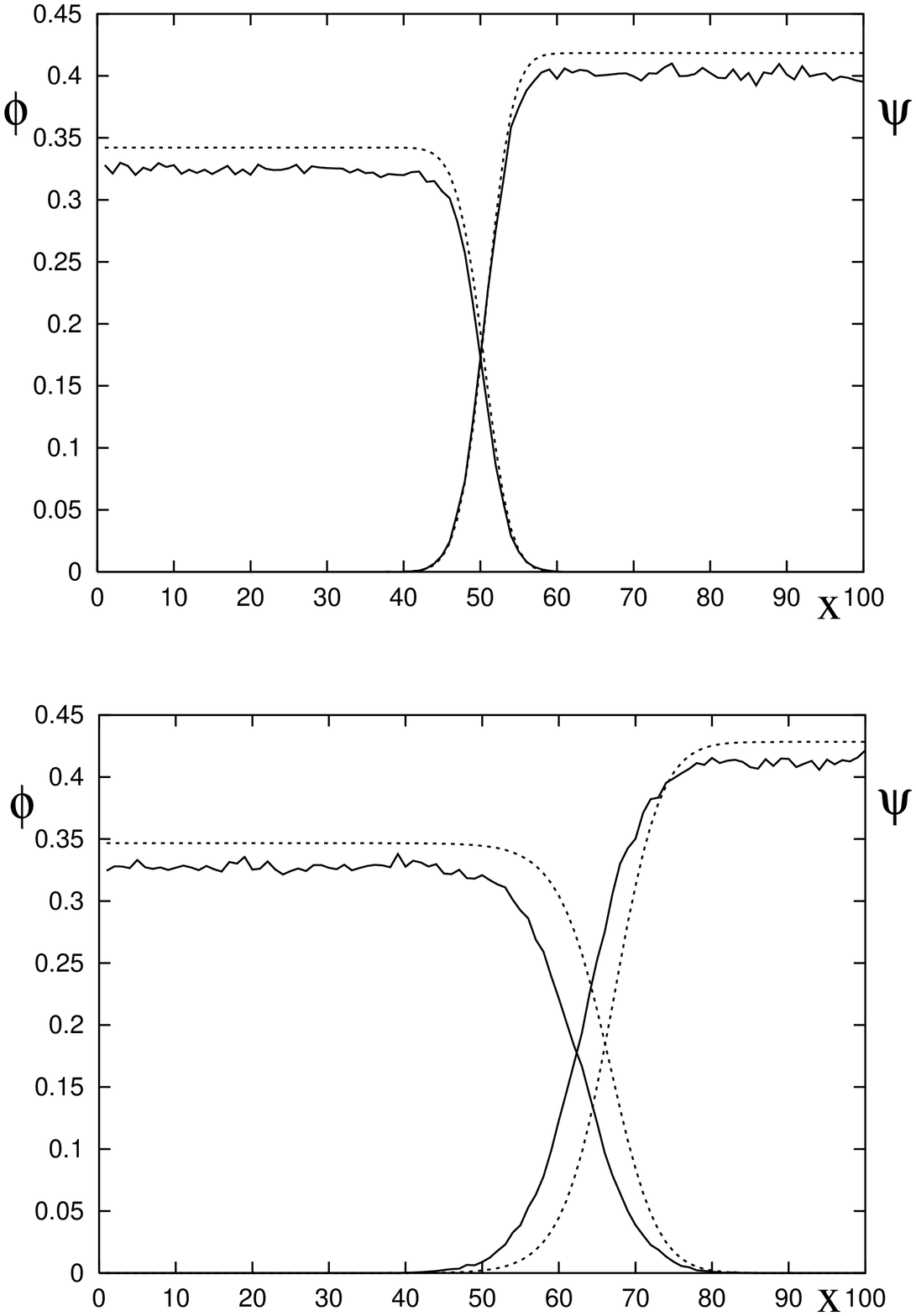} 
\caption{}
\end{figure}

\begin{figure}
\includegraphics[width=5.5in]{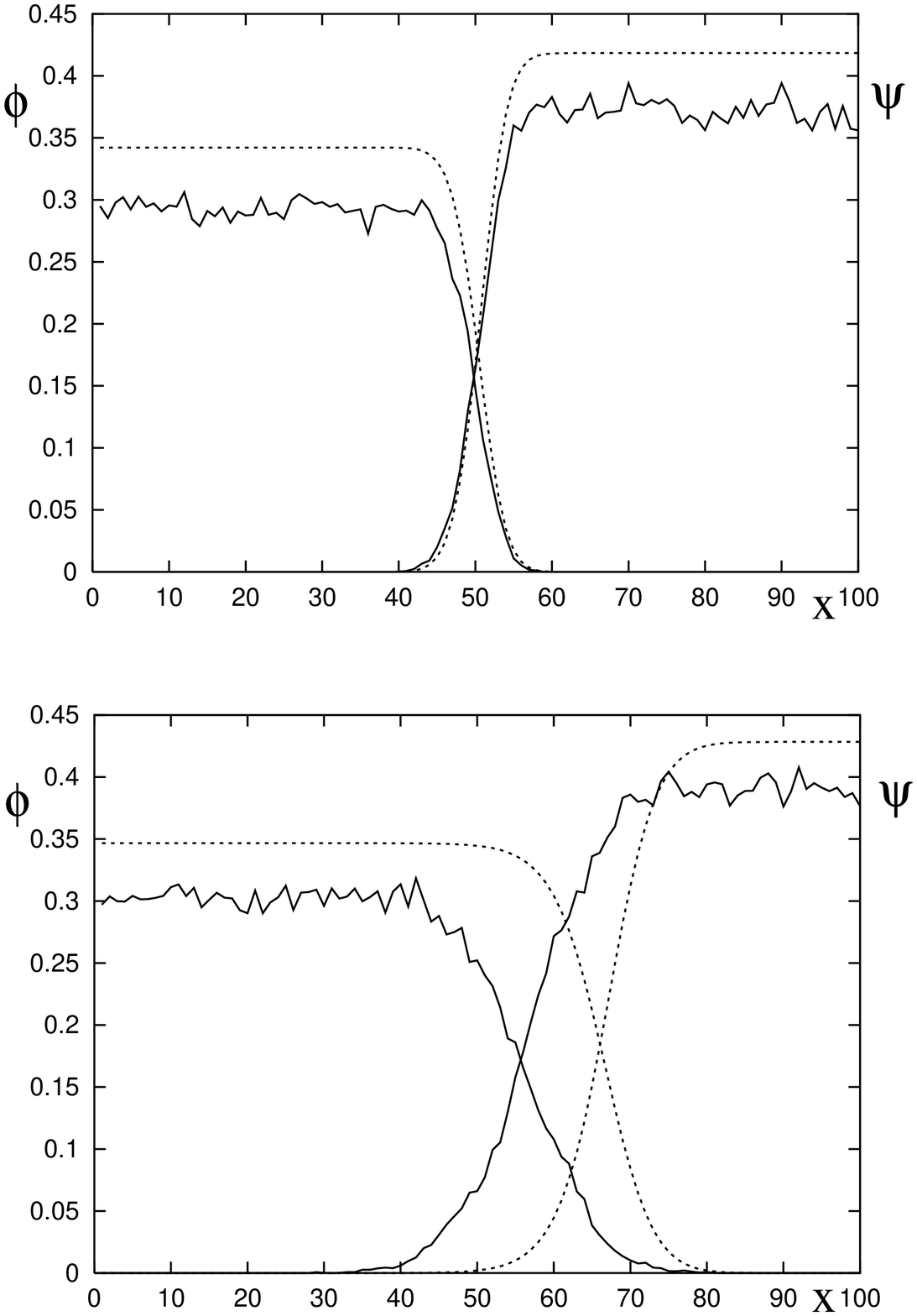} 
\caption{}
\end{figure}

\end{document}